\documentclass[journal=jacsat,manuscript=article]{achemso}

\usepackage[version=3]{mhchem} 

\author{Saeko Tachikawa}
\affiliation[IIS]
{Institute of Industrial Science, The University of Tokyo, Tokyo 153-8505, Japan}
\alsoaffiliation[RCAST]
{Research Center for Advanced Science and Technology, The University of Tokyo, Tokyo 153-8505, Japan}
\email{saekotac@iis.u-tokyo.ac.jp}

\author{Jose Ordonez-Miranda}
\affiliation[IIS]
{Institute of Industrial Science, The University of Tokyo, Tokyo 153-8505, Japan}
\alsoaffiliation[LIMMS]
{LIMMS, CNRS-IIS IRL 2820, The University of Tokyo, Tokyo 153-8505, Japan}
\author{Laurent Jalabert}
\affiliation[IIS]
{Institute of Industrial Science, The University of Tokyo, Tokyo 153-8505, Japan}
\alsoaffiliation[LIMMS]
{LIMMS, CNRS-IIS IRL 2820, The University of Tokyo, Tokyo 153-8505, Japan}

\author{Yunhui Wu}
\affiliation[IIS]
{Institute of Industrial Science, The University of Tokyo, Tokyo 153-8505, Japan}

\author{Yangyu Guo}
\affiliation[IIS]
{Institute of Industrial Science, The University of Tokyo, Tokyo 153-8505, Japan}

\author{Roman Anufriev}
\affiliation[IIS]
{Institute of Industrial Science, The University of Tokyo, Tokyo 153-8505, Japan}

\author{Byunggi Kim}
\affiliation[IIS]
{Institute of Industrial Science, The University of Tokyo, Tokyo 153-8505, Japan}

\author{Hiroyuki Fujita}
\affiliation[IIS]
{Institute of Industrial Science, The University of Tokyo, Tokyo 153-8505, Japan}

\author{Sebastian Volz}
\affiliation[IIS]
{Institute of Industrial Science, The University of Tokyo, Tokyo 153-8505, Japan}
\alsoaffiliation[LIMMS]
{LIMMS, CNRS-IIS IRL 2820, The University of Tokyo, Tokyo 153-8505, Japan}

\author{Masahiro Nomura}
\affiliation[IIS]
{Institute of Industrial Science, The University of Tokyo, Tokyo 153-8505, Japan}
\alsoaffiliation[RCAST]
{Research Center for Advanced Science and Technology, The University of Tokyo, Tokyo 153-8505, Japan}
\alsoaffiliation[LIMMS]
{LIMMS, CNRS-IIS IRL 2820, The University of Tokyo, Tokyo 153-8505, Japan}
\email{nomura@iis.u-tokyo.ac.jp}
\phone{+81 (0)3 5452 6303}

\title[An \textsf{achemso} demo]
  {Polaritonic Waveguide Emits Super-Planckian Thermal Radiation}

\abbreviations{SPhP,Si,SiO$2$}

\begin{document}


\begin{abstract}
Classical Planck’s theory of thermal radiation predicts an upper limit of the heat transfer between two bodies separated by a distance longer than the dominant radiation wavelength (far-field regime).
This limit can be overcome when the dimensions of the absorbent bodies are smaller than the dominant wavelength due to hybrid electromagnetic waves, known as surface phonon-polaritons (SPhPs).
Here, we experimentally demonstrate that the far-field radiative heat transfer between two non-absorbent bodies can also overcome Planck’s limit, by coating them with an absorbent material to form a polaritonic waveguide.
This super-Planckian far-field thermal radiation is confirmed by measuring the radiative thermal conductance between two silicon plates coated with silicon dioxide nanolayers.
The observed conductance is twice higher than Planck’s limit and agrees with the predictions of our model for the SPhP waveguide modes.
Our findings could be applied to thermal management in microelectronics and silicon photonics.
\end{abstract}


Radiative heat transfer between two bodies is an essential phenomenon in our daily lives and technology.
While sunlight powers life on Earth, various other types of radiation play a key role in thermophotovoltaic energy conversion \cite{fan2014alternative, fan2020near, mittapally2021near}, radiative cooling \cite{de2019optical, li2019radiative, zhu2019near, song2020ultrahigh, sadi2020thermophotonic,li2022protecting}, and other applications \cite{yun2017light, bockova2019advances, pacelli2019technological}. 
Classically, the radiative heat transfer is described by Planck's theory \cite{planck1900theory}, which predicts Planck's limit—the maximum radiative heat flux between two macroscopic bodies separated by a distance longer than the dominant radiation wavelength (far-field regime).
But, the radiative flux exceeds this limit under two conditions.
The first condition is the near-field regime when two bodies are closer than the dominant radiation wavelength \cite{polder1971theory, joulain2005surface, kittel2009probing, basu2010near, nomura2016heat, isobe2017spectrally, ito2017dynamic}.
In this regime, the thermal radiation is dominated by the transmission of evanescent electromagnetic waves across the radiating surfaces, separated by a gap of typically nanoscale.
Indeed, experiments showed that radiative flux through a nanogap could exceed Planck's limit by orders of magnitude \cite{rousseau2009radiative, kim2015radiative, song2016radiative, lim2018tailoring, desutter2019near, fiorino2018thermal}.
On the other hand, the second condition allows overcoming the limit in the far-field regime when the dimensions of the radiating body are smaller than the dominant radiation wavelength.
This phenomenon, predicted by theory \cite{biehs2016revisiting, fernandez2018super}, was experimentally demonstrated with suspended silicon nitride nanofilms of a sub-wavelength thickness \cite{thompson2018hundred,thompson2020nanoscale}.
The materials used in the previous works \cite{fernandez2018super, thompson2018hundred, thompson2020nanoscale} are absorbent materials that support in-plane energy transport by surface electromagnetic waves.
These surface waves are generated by the hybridization of photons and optical phonons, and are called surface phonon-polaritons (SPhPs) \cite{chen2005surface, ordonez2013anomalous, lim2019thermal, shin2019far}.
In suspended nanofilms, SPhPs can propagate as far as a few millimeters along the surface, which makes SPhPs powerful energy carriers that could conduct even more heat than phonons \cite{chen2005surface, ordonez2013anomalous, lim2019thermal}. 
Experimental works demonstrated how SPhPs could enhance the in-plane heat conduction of nanofilms made of silicon dioxide \cite{tranchant2019two} and silicon nitride \cite{wu2020enhanced}.
Thus, the in-plane energy of SPhPs, emitted by one body and absorbed by another, could contribute to the radiative heat transfer over Planck's limit, even in the far-field regime.
However, in this work, we aim to demonstrate super-Planckian far-field radiative heat transfer betweem polaritonic waveguides. 
The SPhP waveguides consist of non-absorbent bodies by coating them with absorbent material.
Our experiment is designed to measure far-field radiative heat transfer between two silicon (Si) plates sandwiched by silicon dioxide (SiO$_2$) nanolayers that excite SPhPs.
By comparing the results of this experiment with that of a reference sample without the SiO$_2$ nanolayers, we can confirm the role of SPhPs in radiative heat transfer.

\begin{figure}[ht]
\centering
\includegraphics[width=\linewidth]{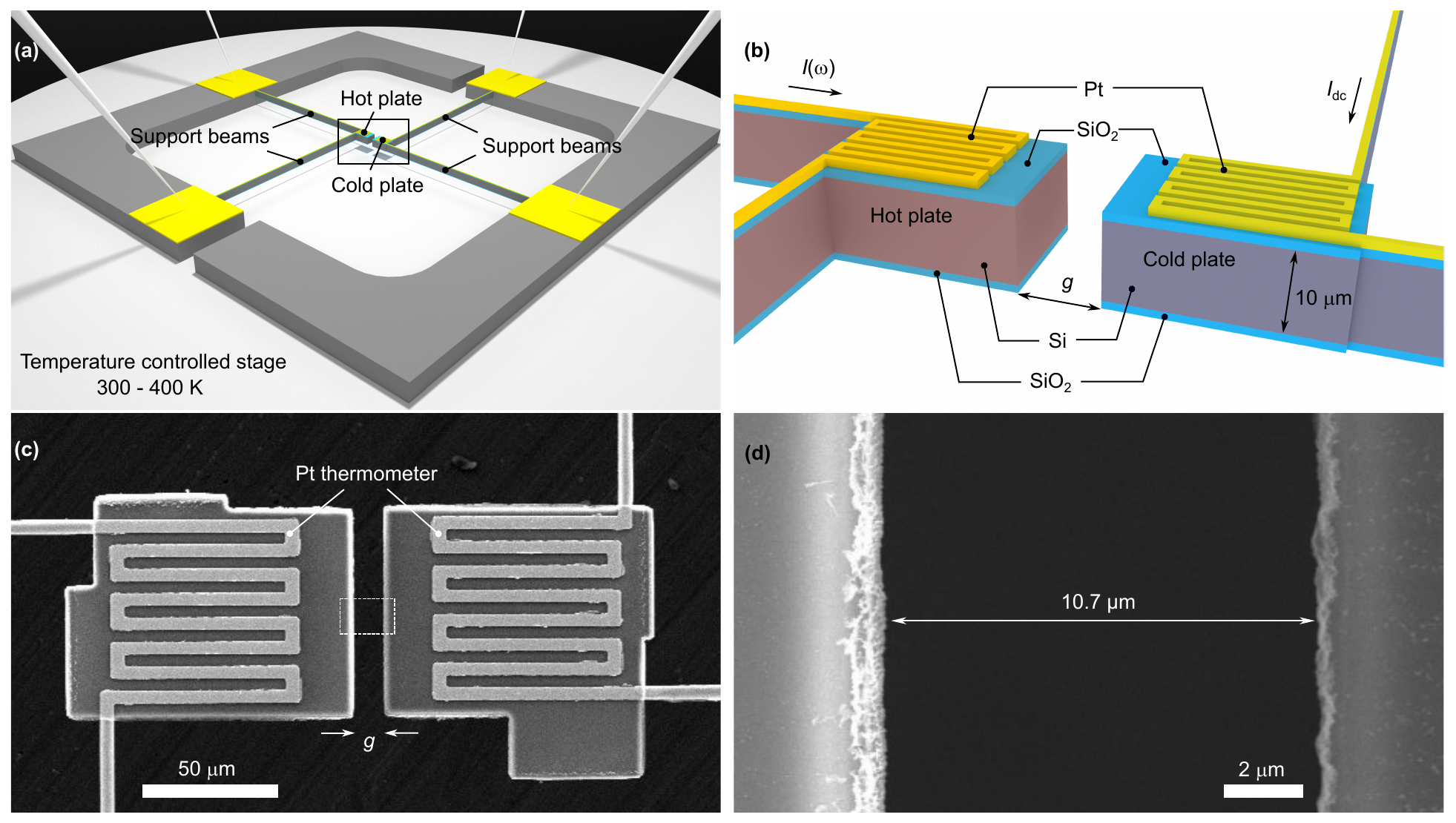}
\caption{\textbf{Experimental platform to measure radiative heat transfer.} (\textbf{a}) Schematic of the device. 
The device with suspended Si plates by support beams is mounted on a temperature-controlled stage, which can raise the temperature from 300 to 400 K. The length of the support beams is 706 {\textmu}m.
(\textbf{b}) Schematic of the hot plate and the cold plate.
The hot and the cold plates are separated by a vacuum gap $g$.
To study the SPhP contribution, the Si plates are sandwiched with SiO$_2$ nanolayers, which generate SPhPs.
Both plates are integrated with Pt thermometers on top.
A sinusoidal electrical current $I(\omega)$, modulated at the frequency $\omega$, circulates along the Pt thermometer of the hot plate, resulting in its temperature rise due to Joule heating.
The radiative heat flux generates the modulated temperature rise on the cold plate, which is calculated from the voltage fluctuation at 2$\omega$ measured by flowing a d.c. electrical current $I_{dc}$ along the thermometer.
Scanning electron microscope (SEM) images showing (\textbf{c}) the top view of the device with the Pt thermometers, and (\textbf{d}) the zoom-in on the region marked with the dotted rectangular box in (\textbf{c}). The actual size of the gap is measured to be $g=$10.7 {\textmu}m.}
\label{Figure1.pdf}
\end{figure}

We fabricated four samples to experimentally quantify the radiative heat transfer between two bodies (Methods).
Each sample consists of two suspended 10-{\textmu}m-thick Si plates separated by a gap, as shown in Fig. \ref{Figure1.pdf}a.
One of the plates (hot plate) was heated up, and the other one (cold plate) experienced a temperature rise caused by absorbing the thermal radiation emitted from the hot plate (Fig. \ref{Figure1.pdf}b).
The temperature of both plates were measured with platinum (Pt) thermometers (Fig. \ref{Figure1.pdf}c).
The separation gap $g$ was 10.7 {\textmu}m to be in the far-field regime (Fig. \ref{Figure1.pdf}d) (Supplementary section 1).
To detect the SPhPs contribution to the radiative heat transfer, we kept one sample as a reference and sandwiched three other ones with SiO$_2$ nanolayers of different thicknesses (30 and 70 nm), as shown in Fig. \ref{Figure1.pdf}b.
Since SiO$_2$ generates SPhPs whereas Si does not, the Si plates with the SiO$_2$ nanolayers are expected to exchange thermal energy via SPhPs, in addition to the conventional thermal radiation.
All samples were placed in a vacuum chamber at a pressure of 10$^{-3}$ Pa to suppress the heat dissipation via the convection \cite{ito2014experimental}.

We measured the temperatures of the plates by using the standard technique of $3\omega$ method \cite{song2015enhancement, song2016radiative,thompson2018hundred, shin2019far} (Methods and Supplementary section 2).
A sinusoidal electrical current $I(\omega)$, modulated with an angular frequency $\omega$, circulates along the Pt wire of the hot plate to cause the temperature rise at an oscillating frequency $2\omega$, $\Delta T_\text{h}$($2\omega$), due to Joule heating.
The thermal radiation from the hot plate resulted in the temperature rise on the cold plate, $\Delta T_\text{c}$($2\omega$), which was calculated from the voltage fluctuation at $2\omega$ measured by flowing a d.c. electrical current along the Pt wire on the cold plate.
Effective gap thermal conductance $G_g$ was calculated by using the d.c. component of the total power generated on the hot plate electrical resistance ($P_{dc}$), $\Delta T_\text{h}$($2\omega$) and $\Delta T_\text{c}$($2\omega$).
Due to the energy conservation upon the input power on the heater, heat flux flowing from the heater to the sensor, and the losses through the supporting beams and radiation, one can solve the thermal circuit equations as below:

\begin{equation}
   P_\text{dc} = G_\text{b}(T_\text{h} - T_{0}) + G_\text{g}(T_\text{h} - T_\text{c}) +  G_\text{loss}(T_\text{h} - T_{0}){,}
\end{equation}
\begin{equation}
   G_\text{g}(T_\text{h} - T_\text{c}) = G_\text{b}(T_\text{c} - T_{0}) + G_\text{loss}(T_\text{c} - T_{0}){,}
\end{equation}
where $T_0$ is ambient temperature, and $G_\text{b}$, $G_\text{g}$ and $G_\text{loss}$ are thermal conductances of the beams, through the gap and of radiation losses, respectively. 
The gap thermal conductance can be obtained as

\begin{equation}
\label{eq:Gg}
G_{\text{g}}=P_{\text{dc}}\frac{\Delta T_{\text{c}}}{(\Delta T_{\text{h}}^2 - \Delta T_{\text{c}}^2)}{,}
\end{equation}
where $\Delta T_\text{h}=T_h-T_0$ and $\Delta T_\text{c}=T_c-T_0$.
The thermal conduction through the beams and radiation losses $G_\text{b} + G_\text{loss}$ were about 18.9, 23.5 and 27.7 {\textmu}W/K for the Si-only system, the three-layer system with the SiO$_2$ layer thickness of 30, and 70 nm, respectively, at the room temperature.

Taking into account that SiO$_2$ is an SPhPs active material whereas Si is not, we determined the SPhP contribution to the radiative heat transfer by comparing $G_g$, with and without the SiO$_2$ nanolayers.
Heat losses by radiation and conduction through the support beams are considered in the calculation, and the measurement setup was sensitive enough to detect radiative heat transfer over the losses.

\begin{figure}[ht]
\centering
\includegraphics[width=\linewidth]{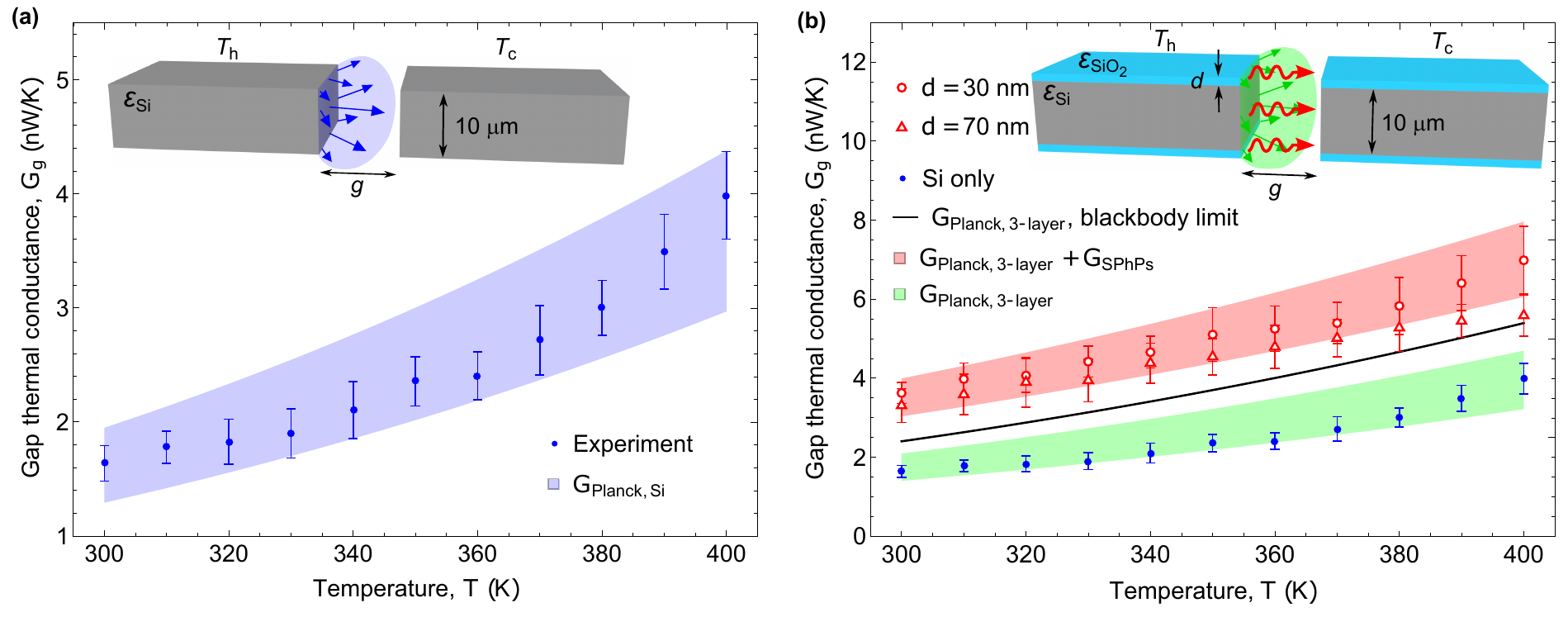}
\caption{ \textbf{The thermal conductance through the gap of experiments and theoretical calculations. (a)} Measured thermal conductance through the gap between two Si plates (Si-only system) and \textbf{(b)} two Si plates sandwiched with SiO$_2$ nanolayers (three-layer system), in a temperature range of 300 to 400 K.
The blue region in \textbf{(a)} represents the prediction of Planck's theory for the far-field thermal conductance $G_\text{Planck, Si}$ determined with a Si emissivity varying from $\epsilon_\text{Si}$ = 0.7 to 0.9.
The green and red zones in \textbf{(b)} show the theoretical predictions of the Planck’s theory for the three-layer system ($G_\text{Planck, 3-layer}$) and Eq. \ref{eq:G_total} for the gap thermal conductance of the three-layer system ($G_\text{Planck, 3-layer}$+$G_\text{SPhP}$), respectively.
Calculation of $G_\text{Planck, 3-layer}$ was done for $\epsilon_\text{Si}$ = 0.7 -- 0.9 and $\varepsilon_{\text{SiO}_{\text2}}$ = 1 to estimate the maximum possible conductance predicted by Planck’s far-field radiation theory. 
For reference, the blackbody limit ($\epsilon_\text{Si}$ = $\varepsilon_{\text{SiO}_{\text2}}$ = 1) is also shown via the black line.
The view factor used in the calculation of $G_\text{SPhP}$ was $F$ = 0.9 -- 1.0 to be appropriate for the in-plane propagation of SPhPs.
The error bars represent
the standard deviations of multiple measurements with different input current $I(\omega)$ and frequencies $\omega$.}
\label{Figure2.pdf}
\end{figure}

First, we benchmarked Planck's radiative heat transfer without SPhPs contribution by measuring the thermal conductance through the gap between Si plates without the SiO$_2$ nanolayers (Si-only system).
Figure \ref{Figure2.pdf}a shows that the measured thermal conductance increases with the stage temperature $T$.
The measured values agree with the prediction of Planck's theory of far-field thermal radiation (Supplementary section 3).
This agreement validates our measurement and indicates that the radiative heat transfer between the Si plates separated by the 10-{\textmu}m-gap can be described by the classical theory.

Figure \ref{Figure2.pdf}b shows the gap thermal conductance between the Si plates sandwiched by the SiO$_2$ nanolayers (three-layer system).
Like in the Si-only system, the measured conductance increases with temperature, as established by the Bose-Einstein distribution function.
However, the measured values of the three-layer systems (open red marks) are nearly twice higher than those obtained for the Si-only system (solid blue circles).
No significant difference was observed between 30 and 70-nm-thick of the SiO$_2$ nanolayers.
We calculated the thermal conductance of the three-layer system based on Planck's far-field theory and plotted it as the green zone. The conductance was calculated with an emissivity of 1 for the SiO$_2$ layers ($G_\text{Planck, 3-layer}$) to estimate the maximum possible heat exchange between the plates (Supplementary section 4).
The calculated conductance of the three-layer system, $G_\text{Planck, 3-layer}$, is slightly higher than that of the Si-only system, $G_\text{Planck, Si}$.
The negligible discrepancy between $G_\text{Planck, 3-layer}$ and $G_\text{Planck, Si}$ is due to the relatively small radiating areas of the SiO$_2$ nanolayers, which are 140 and 300 times thinner than the Si layer.
For comparison, we also calculated and plotted the blackbody limit ($\epsilon_\text{Si}$ = $\varepsilon_{\text{SiO}_{\text2}}$ = 1), as done in the previous works \cite{fernandez2018super, thompson2018hundred}.
Note that the measured conductance $G_g$ of the three-layer system is higher than not only $G_\text{Planck, 3-layer}$, but also the blackbody limit, which establishes that the experimental values of $G_g$ cannot be explained by the classical Planck's theory of far-field radiation.

\begin{figure}[ht]
\centering
\includegraphics[width=\linewidth]{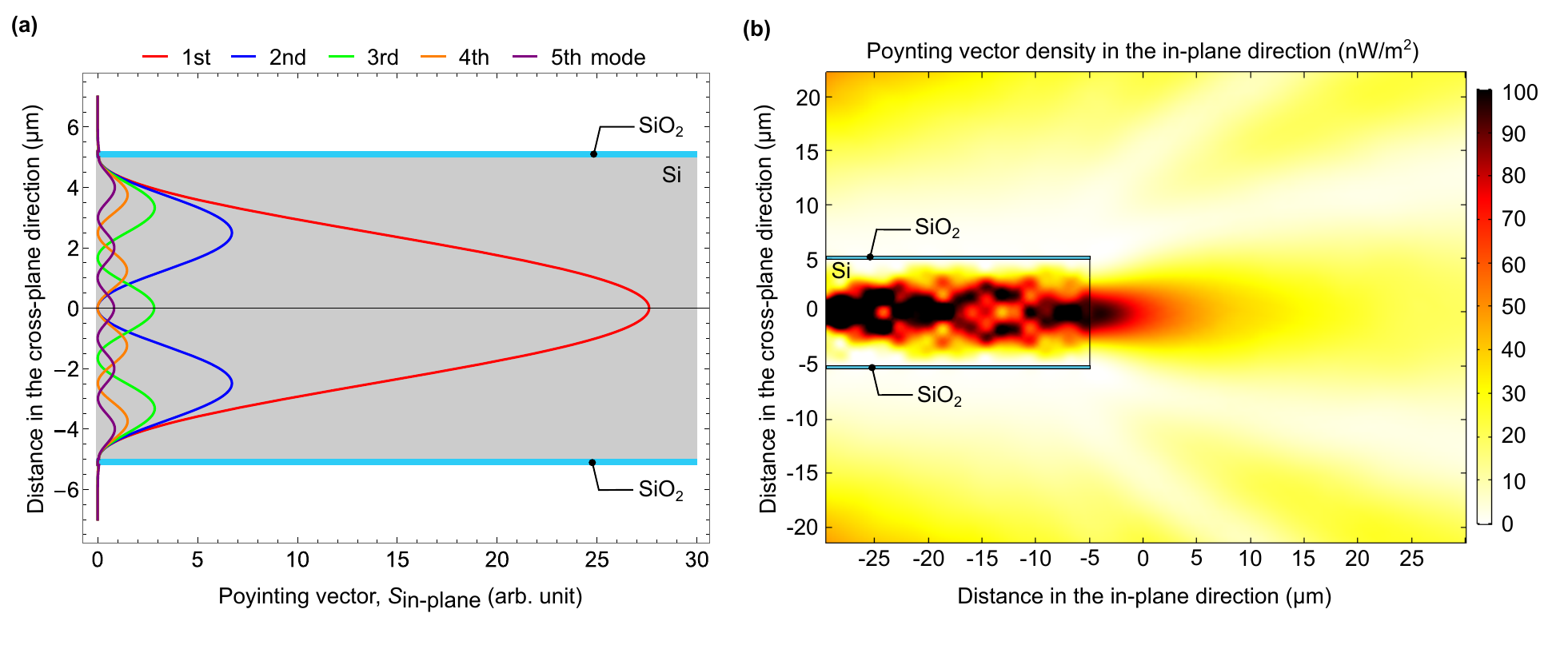}
\caption{\textbf{Energy distribution of SPhPs. (a)}, In-plane Poynting vector inside the three-layer system and \textbf{(b)} its density mapping.
The SPhPs generated in the SiO$_2$ nanolayers hybridize with guided resonant modes inside the non-absorbent Si layer.
The first mode exhibits the largest amplitude, which decreases as the mode order increases.
The density in \textbf{(b)} shows how the electromagnetic flux is emitted from the hot plate along the in-plane direction mainly. 
The calculations were done for the frequency $\omega$ = 203 Trad/s, at which the SPhPs have the strongest cross-plane confinement.}
\label{Figure3.pdf}
\end{figure}

To understand the origin of the increase in gap conductance, we analyzed the spatial energy distribution of the in-plane modes of SPhPs.
The in-plane component of the Poynting vector inside the three-layer system and in vacuum were obtained by solving the Maxwell equations and plotted in Fig. \ref{Figure3.pdf}a and  \ref{Figure3.pdf}b, respectively.
Figure \ref{Figure3.pdf}a shows that the electromagnetic flux is mainly propagating inside the non-absorbent Si layer through guided resonant modes hybridized with SPhPs (Supplementary section 5).
The first mode exhibits the largest amplitude, which decreases as the mode order increases. 
The Si layer thus acts as an SPhP waveguide and the flux is emitted into the vacuum mainly along the in-plane direction, as shown in Fig. \ref{Figure3.pdf}b.
The influence of the Pt thermometer on the in-plane Poynting vector is negligible, as discussed in Supplementary section 7.

Next, we calculated the thermal conductance through the gap to explain why the measured value of the three-layer system was twice higher than that of the Si-only system.
The thermal radiation through the gap consists not only of the far-field radiation between the facing surfaces but also of the radiation caused by SPhPs, which yields the following total thermal conductance through the gap.

\begin{equation}
\label{eq:G_total}
G = G_\text{Planck, 3-layer} + G_\text{SPhPs}{,}
\end{equation}
where $G_\text{SPhPs}$ is the thermal conductance related to SPhPs and is given by 

\begin{equation}
\label{eq:G_SPhPs}
G_\text{SPhPs}=\frac{\epsilon \sigma_\text{2D} a  (T_\text{h}^3 - T_\text{c}^3)}{T_\text{h}-T_\text{c}}{,}
\end{equation}
$\epsilon = \epsilon_\text{3-layer}/(F^{-1}+1-\epsilon_\text{3-layer})$, $\sigma_\text{2D} = 4\zeta (3)k_B^3/ch^2 \approx 9.6 \times 10^{-11}$ W m$^{-1}$K$^{-3}$, $a = 78$ {\textmu}m is the width of the plate, $\epsilon_\text{3-layer}$ is the effective emissivity of the three-layer system, $F$ is the corresponding view factor, $\zeta (x)$ is the zeta function, and $k_B$ and $h$ are the Boltzmann and the Planck constants, respectively.
Considering the relatively large radiation area of the Si layer compared to that of the SiO$_2$ nanolayers, we assumed $\epsilon_\text{3-layer} \approx \epsilon_\text{Si}$ = 0.7 -- 0.9 to estimate $G_\text{SPhPs}$.
Furthermore, since SPhPs propagate in the in-plane direction as shown in Fig. \ref{Figure3.pdf}, Eq. \ref{eq:G_SPhPs} was derived by using the two-dimensional density of states and was evaluated with a view factor of $F$ = 0.9 -- 1.0 (Supplementary section 8).
The obtained prediction of Eq. \ref{eq:G_total} for the total thermal conductance $G$ is shown in Fig. \ref{Figure2.pdf}b through the red zone, which embraces well the corresponding experimental values measured for the three thicknesses of the SiO$_2$ nanolayers.
This fact thus indicates that SPhP waveguide modes in the three-layer system provide an efficient channel to transfer thermal energy along the in-plane direction and overcome the limit of the classical Planck's theory.
This is further confirmed by the fact that the model in Eq. \ref{eq:G_total} and Eq. \ref{eq:G_SPhPs} also predicts well the experimental data reported in the literature for the far-field radiation between two silicon nitride nanofilms \cite{thompson2018hundred}, as detailed in the Supplementary section 9.

In summary, we have experimentally demonstrated the super-Planckian far-field thermal radiation by exciting SPhP waveguide modes.
Using the 3$\omega$ method, we have measured the radiative thermal conductance between two Si plates and found that its value is twice higher when the plates are sandwiched with SiO$_2$ nanolayers.
The conductance increase has been explained by SPhPs, excited in the absorbent SiO$_2$ nanolayers.
The SPhPs form waveguide modes that propagate mainly inside the non-absorbent Si layer, resulting in efficient energy transport along the in-plane direction.
In contrast to the previous studies working with sub-wavelength structures of absorbent materials \cite{song2016radiative, thompson2018hundred, thompson2020nanoscale}, this finding uncovers that SPhPs can enhance the far-field radiation beyond Planck's limit with non-absorbent bodies of dimensions comparable to or greater than the dominant radiation wavelength.
This super-Planckian heat transfer can be obtained using Si-based materials with relatively simple geometry, and therefore, our findings could have broad potential applications in semiconductor fields.

\section{Methods}

\subsection*{Sample preparation}
The devices are fabricated from a silicon-on-insulator wafer, having a top layer of 10 {\textmu}m in thickness, a buried SiO$_2$ layer of 2 {\textmu}m, and a handle layer of 300 {\textmu}m.
First, a backside etching was performed by deep reactive ion etching using the aluminium mask patterned on the backside by photolithography. The buried SiO$_2$ layer worked as an etch stop.
The wafer was cleaned, and the buried SiO$_2$ layer was removed in buffered hydrofluoric acid.
Except for the reference sample, thin SiO$_2$ layers were grown by dry thermal oxidations for the thickness of 30 nm (at 850$^\circ$C for 90 min) and 70 nm (at 1000$^\circ$C for 90 min), on both sides of the top Si layer.
The chromium/platinum (10 nm/100 nm) micro-resistances were patterned on the top Si layer, by using sputtering deposition and lift-off. The top Si structures (beams and micro-resistances platforms) were defined by front-side plasma etching to remove the thermal SiO$_2$, the 10-{\textmu}m-Si layer, and the bottom thermal SiO$_2$.
The remaining photoresist mask was successively removed by oxygen plasma etching as a final step.

\subsection*{Temperature measurement technique}
The heater temperature rise $\Delta T_\text{h}$($2\omega$) is obtained from the third harmonic voltage $3\omega$  measured with a lock-in amplifier in a half-bridge configuration.
The thermal waves, propagating in-plane at $2\omega$ through the vacuum gap, reach the sensor side and are detected with another lock-in amplifier in a d.c. Wheatstone bridge.
The temperature coefficients of resistance for Pt heater and sensor are calculated from the recorded data at the lowest input power to prevent the Joule effect.
All analogue resistances have temperature coefficients of resistance below 5 ppm/K, especially the programmable resistance boxes. 
A Labview program fully controls a matrix of about 300 different experiments resulting from the Cartesian products of four parameters (stage temperature, heater current amplitude and frequency, and sensor d.c. current). 
\begin{acknowledgement}
The authors thank Prof. Hiroshi Toshiyoshi for letting us use his clean room facility.
This work is supported by CREST JST (Grant Number JPMJCR19Q3 and JPMJCR19I1), KAKENHI JSPS (Grant Number 21H04635 and JP20J13729) and JSPS Core-to-Core Program (Grant Number: JPJSCCA20190006).

\end{acknowledgement}

\begin{suppinfo}
These supplementary materials to the paper provide the detailed information about the theoretical background and the analysis of the experimental results. This information is expected to support our discussions and assist the readers for a better understanding of our results. Further discussion is welcomed by contacting our corresponding authors.
\end{suppinfo}

\bibliography{Reference}

\end{document}


\maketitle

\begin{abstract}
\textbf{
These supplementary materials to the paper provide the detailed information about the theoretical background and the analysis of the experimental results. This information is expected to support our discussions and assist the readers for a better understanding of our results. Further discussion is welcomed by contacting our corresponding authors.  
} 
\end{abstract}

\section*{Supplementary}
\section{1. Evaluation of near-field radiative heat transfer contribution to the measured gap conductance}

\begin{figure}[ht]
\centering
\includegraphics[width=0.5\linewidth]{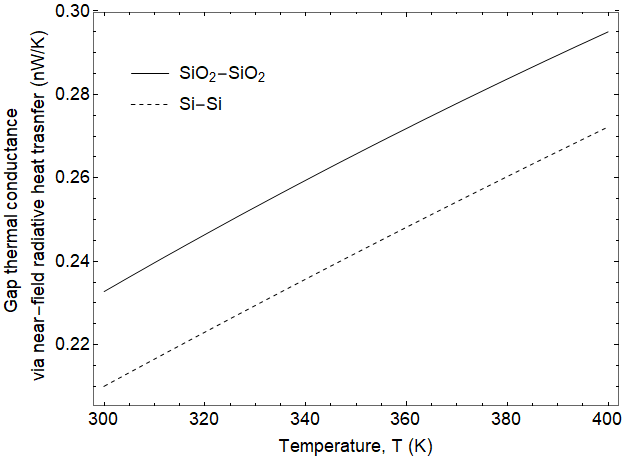}
\caption{$\mid$ \textbf{The gap thermal conductance by near-field radiative heat transfer.} The conductance was calculated between infinite Si plates and between infinite SiO$_2$ plates, separated by a vacuum gap of 10 {\textmu}m}
\label{NFRHT.png}
\end{figure}

To evaluate the contribution of near-field effects to the measured $G_{g}$, we estimated the near-field radiative heat transfer between infinite Si plates and between infinite SiO$_2$ plates, separated by a vacuum gap of 10 {\textmu}m.
The heat transfer coefficient per unit area is \cite{polder1971theory,mulet2002enhanced}:

\begin{equation}
   h_\text{NF} = \sum_{i = s,p} \int d\omega \int^{\infty}_{\frac{\omega}{c}} \frac{k}{2\pi^3} dk \frac{\partial}{\partial T} \left(\frac{\hbar\omega}{e^{\frac{\hbar\omega}{k_BT}}-1}\right) \frac{\mathrm{Im}[r^i_{31}] \mathrm{Im}[r^i_{32}] e^{-2\mathrm{Im}[\gamma_3]d}}{\left| 1-r^i_{31}r^i_{32} e^{2i\gamma_3d} \right|^2}{,}
\end{equation}
where $s,p$ correspond to s- and p-polarization, 1,2 and 3 are for emitter, receiver and vacuum in between, respectively. Further, $\gamma_3$ is a cross-plane wave vector in vacuum, $r^i_{31,32}$ is Fresnel reflection factors.
The gap thermal conductance for evanescent wave was calculated by the heat transfer coefficient derived above multiplied by the actual cross-section area for our device (Fig. \ref{NFRHT.png}).
Note that in both cases, these upper bounds of the near-field contribution are more than one order of magnitude lower than $G_\text{g}$, which confirms that the measured values are determined by the far-field radiative heat transfer.
Since it was calculated between infinite plates, actual near-field effects in our experiments can be predicted even lower, therefore negligible.

\section{2. Experimental setup based on 3$\omega$ method}

\begin{figure}[ht]
\centering
\includegraphics[width=0.7\linewidth]{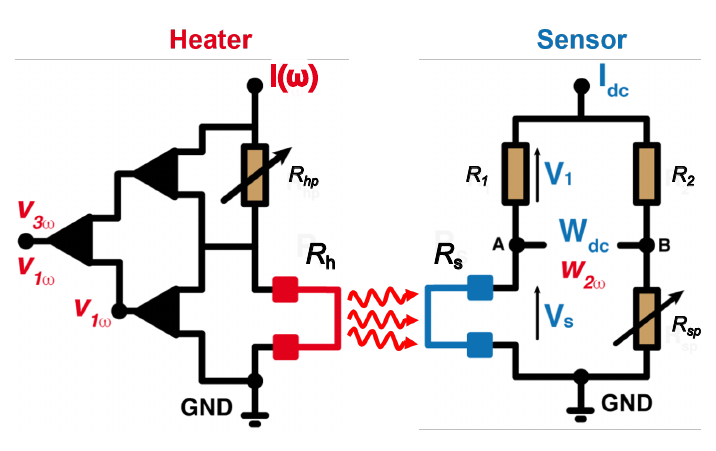}
\caption{$\mid$ \textbf{Electrical circuit schematics of our setup based on the 3$\omega$ method.} The heater is connected to a half-bridge with a fluctuating input and the sensor is integrated in a d.c. Wheatstone bridge, to filter 3$\omega$ and 2$\omega$ signal respectively.}
\label{3omegasetup.pdf}
\end{figure}

The setup configuration is shown in Fig. \ref{3omegasetup.pdf}.
The heater micro-resistance is connected to a classic AC half-bridge and in series with a programmable resistance decade box $R_{\text{hp}}$.
The third amplifier, taking the differential voltage between $R_\text{hp}$ and $R_\text{h}$, intends to cancel the d.c. signals.
A dual-frequency lock-in amplifier (LIA) at the output of the third differential amplifier (AD524), is measuring both the residual 1$\omega$ voltage between $R_\text{hp}$ and $R_\text{h}$, and the 3$\omega$ voltage. 
The bridge is balanced at 300 K and for the highest frequency in order to minimise the Joule effect on $R_{\text{h}}$. A low noise AC current source (Lakeshore AC155) supplies the current $I(\omega)$ in both resistances, $R_\text{hp}$ and $R_\text{h}$. 

The thermal waves emitted at 2$\omega$ from the hot plate travel through the 10 {\textmu}m vacuum gap and reach the cold plate.
We use a d.c. Wheatstone bridge connected to a low noise d.c. current source (Lakeshore AC155) for the sensor side.
The resistance $R_1$ and $R_2$ are 1 k$\Omega$ each. The $R_{\text{sp}}$ is a programmable decade resistance box used for balancing the bridge at a stage temperature of 300 K, and for the highest heater frequency (700 Hz), i.e. when the thermal radiation from the heater is negligible.
We measure the d.c. voltages on $R_1$ and $R_{\text{s}}$ with 6.5 digits multimeters (Agilent 34410A and Keysight 34461A). A nano voltmeter (Keithley 2182) monitors the residual voltage in the bridge, in parallel with an LIA (SR 830) measuring the 2$\omega$ voltage induced by the heat flux from the heater side. 
The resistances $R_1$, $R_2$, $R_{\text{hp}}$ and $R_{\text{sp}}$ have temperature coefficient of resistance (TCR) below 5 ppm/K.

The device is loaded in a vacuum probe station, with the stage temperature ranging from 300 K to 400 K, and controlled with a temperature controller (Lakeshore 335). 

A Labview program is driving the stage temperature, the heater current amplitude and frequency, and the sensor current amplitude, in a full automated way, including a self-adjustment of the LIA sensitivities according to the input signal.  

The 3$\omega$ method is a method in which AC modulated input generates modulated temperature, thus electrical signals also modulating at a certain frequency can be filtered to achieve high sensitivity.
The heater and sensor micro-resistances are measured with a 2-wires configuration. The classical 3$\omega$ equations have been revisited to take into account the possible non-linear TCR.
On the heater side, the injected power $P_\text{input}$ on $R_{\text{h}}$ is given by
\begin{equation}
    P_\text{input}=R_\text{h0}I^2=R_\text{h0}I_{0}^2\cos^2{(\omega t)}=P_\text{dc}\left[
1 + \cos{(2\omega t)}\right]
=P_\text{dc}\mathrm{Re}(1+\mathrm{e}^{i2\omega t}){,}
\end{equation}
while the current is $I = I_{0}\cos {(\omega t)}$, $P_\text{dc}=\frac{1}{2} R_\text{h0} I_{0}^2$ is the d.c. power contribution, with $R_\text{h0}=V_{0}/I_{0}$ being the heater resistance measured at $T=T_{0}=$300 K and the highest frequency (=700 Hz) to neglect Joule heating.
The heater voltage amplitude $V_{0}$ is measured at the output of a low noise instrumentation amplifier with a lock-in amplifier.
Assuming a quadratic variation of the resistance with the total temperature $T$, we define the total resistance $R_\text{h}$ as
\begin{equation}
    R_\text{h}(T)=R_\text{h0}\left[
1 + \alpha (T-T_{0}) +  \beta (T-T_{0})^2 + ...  \right]{.}
\end{equation}
with the linear TCR $\alpha=R_\text{h}'(T_0)/R_\text{h}(T_0)$, and the non-linear TCR $\beta=R_\text{h}''(T_0)/2R_\text{h}(T_0)$.
By analogy, the variation of the heater resistance induces a variation of the voltage $V_\text{h}$ given by

\begin{equation}
\label{eq:Vh(T)}
   V_\text{h}=R_\text{h}I=V_{0}\left[
1 + \alpha (T-T_{0}) +  \beta (T-T_{0})^2 + ...  \right]\mathrm{Re}(\mathrm{e}^{i\omega t}){.}
\end{equation}
Assuming that contributions from the temperature with the order higher than the third one are negligible in Eq. \ref{eq:Vh(T)}, the total voltage $V_\text{h}$ can be rewritten introducing the ﬁrst, third and ﬁfth harmonics,

\begin{equation}
\label{eq:Vh}
   V_\text{h} = V_{0}\left[
1 + \alpha (T-T_{0}) +  \beta (T-T_{0})^2 \right]\mathrm{Re}(\mathrm{e}^{i\omega t}) = \mathrm{Re}(V_{\omega}\mathrm{e}^{i\omega t} + V_{3\omega}\mathrm{e}^{3i\omega t} + V_{5\omega}\mathrm{e}^{5i\omega t}){.}
\end{equation}
The total temperature $T$ is the sum of non-modulated temperature $\Delta T_\text{h,dc}$  and the real part Re$(T_{\text{h},2\omega} e^{2i\omega t})$ of the modulated temperature $\Delta T_{\text{h},2\omega}$ due to Joule heating,

\begin{equation}
\label{eq:T}
   T-T_0 = \Delta T_\text{h,dc} + \mathrm{Re}(T_{\text{h},2\omega} \mathrm{e}^{2i\omega t}){.}
\end{equation}
Substituting Eq. \ref{eq:T} into Eq. \ref{eq:Vh}, we obtain the expressions of harmonics voltages as a function of the linear and non-linear TCRs, such as

\begin{equation}
   V_{\omega}=V_{0}\left[1 + \alpha \Delta T_\text{h,dc} +\beta \Delta T_\text{h,dc}^2 + \frac{1}{2} (\alpha+2\beta \Delta T_\text{h,dc})T_{2\omega} + \frac{\beta}{2} |T_{\text{h},2\omega}|^2  \right]{,}
\end{equation}

\begin{equation}
\label{eq:V3w}
   V_{3\omega}=\frac{V_{0}}{2} \left[(\alpha + 2\beta \Delta T_\text{h,dc})T_{\text{h},2\omega} + \frac{\beta}{2} |T_{\text{h},2\omega}|^2  \right]{,}
\end{equation}

\begin{equation}
   V_{5\omega}=\frac{V_{0}}{4} \beta T_{\text{h},2\omega}^2{.}
\end{equation}
The heater temperature rise $T_{\text{h},2\omega}$ could be directly retrieved from the fifth harmonic voltage, and the non-linear TCR coefficient of $\beta$.
However, the fifth harmonic signal amplitude is challenging to measure as it is closed to the noise level of the overall instrumentation.
Therefore we use Eq. \ref{eq:V3w} to retrieve $T_{\text{h},2\omega}$, by assuming a negligible Joule heating so that the heater voltage $V(\omega \rightarrow \infty)$ is the one measured at the highest frequency, which  leads to the expression

\begin{equation}
\label{eq:V1w}
   V_{\omega}(\omega \rightarrow \infty) \equiv V_{\infty} =V_{0}(1 + \alpha \Delta T_\text{h,dc} +\beta \Delta T_\text{h,dc}^2 ){.}
\end{equation}
Here, we introduce the coefficient $S$ as $\alpha+\beta \Delta T_\text{h,dc}=\sqrt{S}$. The coefficient $S$ can be obtained by 

\begin{equation}
\label{eq:S}
   S = \alpha^2 - 4\beta \left
(1- \frac{V_{\infty}}{V_{0}}\right){.}
\end{equation}
The temperature elevation by Joule heating on the heater side $T_{\text{h},2\omega}$ is derived from Eq. \ref{eq:V1w}, Eq. \ref{eq:S}, together with Eq. \ref{eq:V3w}:

\begin{equation}
   \frac{V_{3\omega}}{V_{0}} = \sqrt{S} \frac{T_{\text{h},2\omega}}{2} + \beta \left
(\frac{T_{\text{h},2\omega}}{2}\right)^2{,}
\end{equation}

\begin{equation}
T_{\text{h},2\omega} = \frac{\sqrt{S+4\beta V_{3\omega}/V_{0}}-\sqrt{S}}{\beta}{.}
\end{equation}
We can verify that when the non-linear TCR $\beta \rightarrow 0$, we obtain the classical expression of the $3\omega$ method,

\begin{equation}
   V_{3\omega}=\frac{V_{0}}{2} \alpha |T_{\text{h},2\omega}|^2{,}
\end{equation}

On the sensor side, we use a classic Wheatstone bridge with a d.c. current input $I_\text{dc}$.
By analogy with the heater side, we can write the total sensor voltage $V_\text{s}$ as,

\begin{equation}
  V_\text{s} = R_\text{s}I_\text{dc}=V_\text{s0}\left[
1 + \alpha (T_\text{s}-T_{0}) +  \beta (T_\text{s}-T_{0})^2 + ...  \right]{.}
\end{equation}
and define the total sensor temperature $T_\text{s}$ as $T_\text{s}-T_0 = \Delta T_\text{s,dc} + \mathrm{Re}(T_{\text{s},2\omega} \mathrm{e}^{2i\omega t})$, where $\Delta T_\text{s,dc}$ is the non-modulated (i.e "d.c.") temperature mainly due to the stage temperature, and $T_{\text{s},2\omega}$ is the amplitude of the modulated temperature that is due only to the thermal waves coming from the heater side.
The total sensor voltage is a sum of a d.c. component $V_\text{m}$ and an AC component that contains the second and fourth harmonics, such as

\begin{equation}
   V_\text{s} = V_\text{m} + \mathrm{Re}(V_{2\omega}\mathrm{e}^{2i\omega t} + V_{4\omega}\mathrm{e}^{4i\omega t}){,}
\end{equation}
where

\begin{equation}
   V_\text{m}=V_\text{s0}\left[1 + \alpha \Delta T_\text{s,dc} +\beta \Delta T_\text{s,dc}^2 + \frac{\beta}{2} |T_{\text{s},2\omega}|^2  \right]{,}
\end{equation}

\begin{equation}
   V_{2\omega}=V_\text{s0} \left(\alpha + 2\beta \Delta T_\text{s,dc}\right)T_{\text{s},2\omega}{,}
\end{equation}

\begin{equation}
   V_{4\omega}=\frac{V_\text{s0}}{2} \beta T_{\text{s},2\omega}^2{.}
\end{equation}
Here, the temperature raise $T_{\text{s},2\omega}$ can be directly retrieved from the 4th harmonic signal.
Practically, the $4\omega$ signal amplitude is much smaller than the $2\omega$ one (in the 10 nV -- 10 {\textmu}V range), which is more challenging to measure due to the large gap between the hot and cold plates.
In order to calculate $T_{\text{s},2\omega}$ from the 2nd harmonic signal, we assume (as the heater case) a negligible temperature elevation on the sensor side for the highest heater frequency (700Hz). Under this assumption, we can write the d.c. component $V_\text{m}$ as: 

\begin{equation}
   V_\text{m}(\omega \rightarrow \infty) \equiv V_{\text{s}\infty} =V_\text{s0}(1 + \alpha \Delta T_\text{s,dc} +\beta \Delta T_\text{s,dc}^2 ){.}
\end{equation}
with $V_{s0}$ the sensor voltage at high frequency, and $T=T_0$. Again, by introducing the $S$ coefficient for the sensor side, $\alpha+\beta \Delta T_\text{dc}=\sqrt{S}$, thence $S = \alpha^2 - 4\beta \left
(1- \frac{V_{\text{s}\infty}}{V_\text{s0}}\right)$, the modulated temperature on the sensor $T_{\text{s},2\omega}$ is given by

\begin{equation}
   T_{s,2\omega} = \frac{V_{2\omega}}{\sqrt{S}V_\text{s0}}{.}
\end{equation}
We verify that in case of a linear $R(T)$, the coefficient $\beta$ tends to zero, and $T_{\text{s},2\omega} = V_{2\omega} / (\alpha V_{\text{s}0})$.
From a measurement point of view, $V_{2\omega}$ would be the $2\omega$ voltage measured directly on $R_\text{s}$. To prevent large input signal on the lock-in amplifier, the $2\omega$ signal is rather measured from the differential voltage $W_{2\omega}$ in the bridge in order to reduce the unwanted d.c. signals. The total voltage in the bridge $V_\text{AB}$ is

\begin{equation}
   V_\text{AB} = \frac{R_{2}V_\text{s}-R_\text{sp}V_{1}}{R_{2}+R_\text{sp}}{.}
\end{equation}
Since $V_\text{s}=V_\text{m} + \mathrm{Re}(V_{2\omega}\mathrm{e}^{2i\omega t} + V_{4\omega}\mathrm{e}^{4i\omega t})$, $V_\text{AB}$ can be also described as $V_\text{AB}=W_\text{dc} + \mathrm{Re}(W_{2\omega}\mathrm{e}^{2i\omega t} + W_{4\omega}\mathrm{e}^{4i\omega t})$. 
The $2\omega$ sensor voltage $V_{2\omega}$ can be obtained from the measured bridge voltage $W_{2\omega}$ by the expression

\begin{equation}
   V_{2\omega} =\left(1+ \frac{R_\text{sp}}{R_{2}}\right)W_{2\omega}{.}
\end{equation}
This signal is filtered again by a lock-in amplifier connected between the points A and B.
Finally, the modulated temperature $T_{\text{s},2\omega}$ on the sensor side is given by
\begin{equation}
   T_{s,2\omega} = \left(1+ \frac{R_\text{sp}}{R_{2}}\right)\frac{W_{2\omega}}{\sqrt{S}V_\text{s0}}{.}
\end{equation}

\section{3. Far-field radiative heat transfer between finite bodies}

\begin{figure}[ht]
\centering
\includegraphics[width=\linewidth]{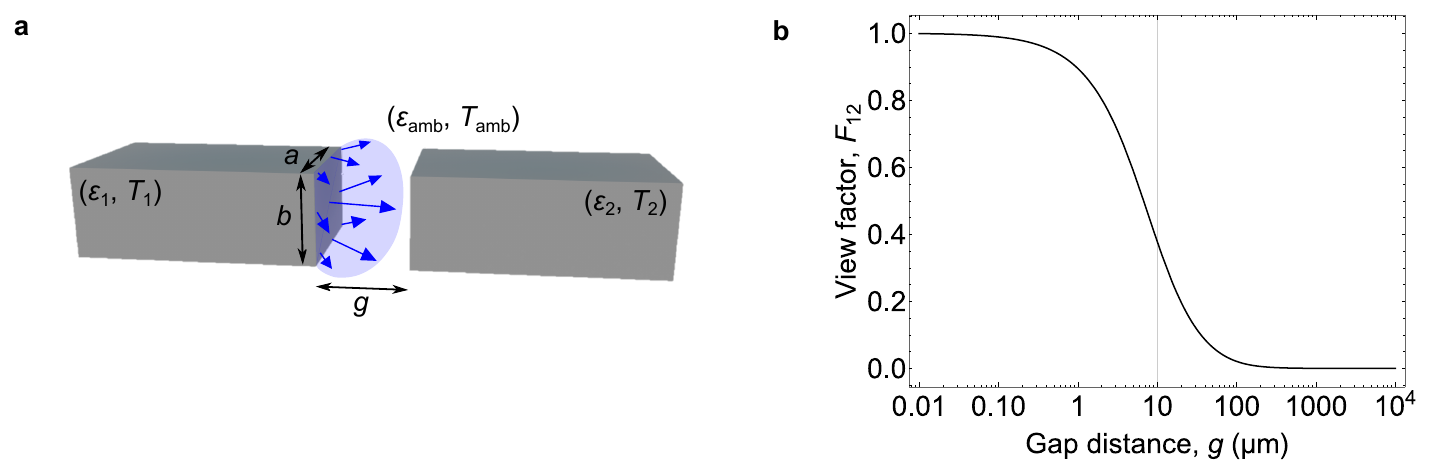}
\caption{$\mid$ \textbf{Far-field radiative heat transfer between finite bodies.} \textbf{a}, Schematic of the emitter and the receiver exchanging heat flux. \textbf{b}, The view factor typically for our device geometry ($a$ = 78 {\textmu}m and $b$ = 10 {\textmu}m) for different gap distance.}
\label{FFRHT.pdf}
\end{figure}

The theory for far-field radiative heat transfer is well-established by Stefan–Boltzmann law. 
It describes the heat flux radiated from a body with the emissivity of $\epsilon$, $q$, proportional to 4th power of the body's temperature,

\begin{equation}
  q = \epsilon \sigma T^4{,}
\end{equation}
where $\sigma$ is the Stefan–Boltzmann constant. The power exchanged between the surface $i$ and $j$ is described as,

\begin{equation}
  q_{ij} = \sigma A_i \epsilon_i F_{ij} (T_i^4-T_j^4) = \sigma A_j \epsilon_j F_{ji} (T_i^4-T_j^4) {,}
\end{equation}
where $F_{ij}$ is a view factor, which is a geometrical function to indicate how much heat flux leaving one surface can reach the other surface. It is given by

\begin{equation}
  F_{ij} = \frac{1}{A_i} \int_{A_i} \int_{A_j} \frac{\cos{\theta_i}\cos{\theta_j}}{\pi r_{ij}^2} dA_i dA_j {,}
  F_{ji} = \frac{1}{A_j} \int_{A_i} \int_{A_j} \frac{\cos{\theta_j}\cos{\theta_i}}{\pi r_{ji}^2} dA_j dA_i {,}
\end{equation}
where $A_{i,j}$ is the surface areas of each surface, $\theta_{i,j}$ is the angles between the surface normals and a ray between the two differential areas. The distance between the differential areas is $r_{ij}$. The relationship $A_iF_{ij}=A_jF_{ji}$ is always satisfied. 
The heat fluxes from N numbers of surfaces can be written in a matrix form as

\begin{equation}
\label{eq:heat flux}
  \sum_{j=1}^{N} \chi_{ij} q_{ij} = \sigma \sum_{j=1}^{N} F_{ij}\left(T_i^4-T_j^4\right){,}
\end{equation}
where $\chi_{ij}=-F_{ij}(\epsilon_j^{-1}-1)$.
Here, we consider 3 surfaces as shown in Fig. \ref{FFRHT.pdf}a, in which the emitter with the emissivity and the temperature of ($\epsilon_1$,$T_1$) and the receiver with ($\epsilon_2$,$T_2$) are positioned in vacuum with ambient temperature of $T_{amb}$, exchanging heat fluxes. The emitter and the receiver are in the same geometry, corresponding with our device so that their opposing surfaces have same areas, $A_1=A_2=ab$. Applying this condition to Equation \ref{eq:heat flux} gives,

\begin{equation}
\label{eq:matrix1}
    \left[
    \begin{array}{ccc}
    \chi_{11} & \chi_{12} & \chi_{13} \\
    \chi_{21} & \chi_{22} & \chi_{23} \\
    \chi_{31} & \chi_{32} & \chi_{33}
    \end{array}
    \right]
    \left[
    \begin{array}{c}
    q_{1}\\
    q_{2}\\
    q_{3}
    \end{array}
    \right]
=\sigma
    \left[
    \begin{array}{c}
    F_{12}\left(T_1^4-T_2^4\right) + F_{13}\left(T_1^4-T_{amb}^4\right)\\
    F_{21}\left(T_2^4-T_1^4\right) + F_{23}\left(T_2^4-T_{amb}^4\right)\\
    F_{31}\left(T_{amb}^4-T_1^4\right) + F_{32}\left(T_{amb}^4-T_2^4\right)
    \end{array}
    \right]{,}
\end{equation}
while $F_{11}=F_{22}$. Since $\sum_{j=1}^{N} F_{ij} =1$ and $A_iF_{ij}=A_jF_{ji}$ are satisfied, the view factors between the device surfaces and the ambient surface can be written as

\begin{equation}
\begin{split}
  &F_{13} = 1-F_{12}{,}\\
  &F_{23} = 1-F_{21}= 1-F_{12}{,}\\
  &F_{33} = 1-F_{31}-F_{32} = 1 - A_{13}F_{13} - A_{23}F_{23} = 1-2A_{13}(1-F_{12}){,}
\end{split}
\end{equation}
where $A_{ij}=\frac{A_j}{A_i}$. The $\chi_{ij}$ in Equation \ref{eq:matrix1} are 

\begin{equation}
\label{eq:matrix_Khi}
    \left[
    \begin{array}{ccc}
    \chi_{11} & \chi_{12} & \chi_{13} \\
    \chi_{21} & \chi_{22} & \chi_{23} \\
    \chi_{31} & \chi_{32} & \chi_{33}
    \end{array}
    \right]
=
    \left[
    \begin{array}{ccc}
    \epsilon_{1}^{-1} & -F_{12}(\epsilon_{2}^{-1}-1) & 0\\
    -F_{12}(\epsilon_{1}^{-1}-1) & \epsilon_{2}^{-1} & 0\\
    -A_{13}(1-F_{12})(\epsilon_{1}^{-1}-1) & -A_{13}(1-F_{12})(\epsilon_{2}^{-1}-1) & 0
    \end{array}
    \right]{,}
\end{equation}
since $\epsilon_{vac}=1$. Equation \ref{eq:matrix_Khi} yields Equation \ref{eq:matrix1} described as

\begin{equation}
\begin{split}
  &q_{1}D = \left[\epsilon_2^{-1} - F_{12}^2(\epsilon_{2}^{-1}-1)\right]\sigma(T_1^4-T_{amb}^4) - \sigma F_{12}(T_2^4-T_{amb}^4){,}\\
  &q_{2}D = \left[\epsilon_1^{-1} - F_{12}^2(\epsilon_{1}^{-1}-1)\right]\sigma(T_2^4-T_{amb}^4) - \sigma F_{12}(T_1^4-T_{amb}^4){,}
\end{split}
\end{equation}
where $D=\epsilon_1^{-1}\epsilon_2^{-1}-F_{12}^2(\epsilon_1^{-1}-1)(\epsilon_2^{-1}-1)$. Since the heat exchange between the emitter and the receiver is $q_{12}=F_{12}\left[\sigma(T_1^4-T_2^4) + (\epsilon_{2}^{-1}-1)q_{2} - (\epsilon_{1}^{-1}-1)q_{1} \right]$, $q_{12}$ can be described as,

\begin{equation}
  q_{12} = \epsilon \sigma \left\{(T_1^4-T_2^4) + \frac{\left(F_{12}^{-1}-1\right)}{\epsilon_1\epsilon_2}\left[\epsilon_1(T_1^4-T_{amb}^4)-\epsilon_2(T_2^4-T_{amb}^4)\right]\right\} {,}
\end{equation}
where $\epsilon^{-1}=\epsilon_1^{-1}+\epsilon_2^{-1}-1+(F_{12}^{-1}-1)\epsilon_1^{-1}\epsilon_2^{-1}$. In our case of the device with only Si layer, $\epsilon_1=\epsilon_2=\epsilon_{Si}$ therefore the above equation can be reduced to 

\begin{equation}
  q_{12} =  \frac{\epsilon_{Si}\sigma \left(T_1^4-T_2^4\right)}{F_{12}^{-1}+1-\epsilon_1} {.}
\end{equation}
One can see that it is independent of $T_{amb}$. 
The gap thermal conductance between the emitter and the receiver via far-field radiative heat transfer (FFRHT) is,

\begin{equation}
  G_{FFRHT} = \frac{q_{12}}{T_1-T_2}{.}
\end{equation}
For the configuration of opposing rectangles, the view factor is defined typically as 

\begin{equation}
\label{eq:viewfactor}
  F_{12} =  \frac{2}{\pi x y} \left[
  \ln{\sqrt{\frac{(1+x^2)(1+y^2)}{1+x^2+y^2}}} + x\sqrt{1+y^2}\tan^{-1}\left(\frac{x}{\sqrt{1+y^2}}\right) + y\sqrt{1+x^2}\tan^{-1}\left(\frac{y}{\sqrt{1+x^2}}\right) -x\tan{x}^{-1} -y\tan{y}^{-1} \right]{,}
\end{equation}
where $x=\frac{a}{g}$ and $y=\frac{b}{g}$. The view factor for our device geometry of $a$ = 78 {\textmu}m and $b$ = 10 {\textmu}m is plotted for different gap distance in Fig. \ref{FFRHT.pdf}b.

\section{4. Planck's classical far-field radiative heat transfer between three-layer bodies}

\begin{figure}[ht]
\centering
\includegraphics[width=0.5\linewidth]{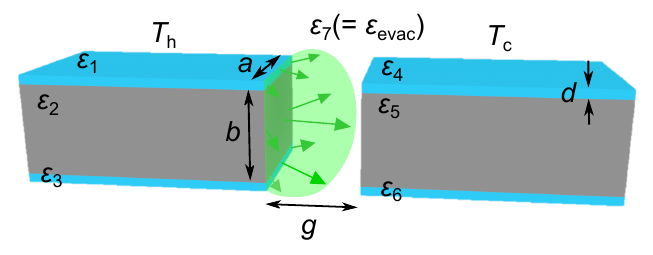}
\caption{$\mid$ \textbf{Planck's conventional far-field radiative heat transfer between three-layer systems.} Schematic of the emitter and the receiver of three-layer systems exchanging heat flux.}
\label{3layerFFRHT.pdf}
\end{figure}

Here, we show the calculation of the far-field radiative heat transfer between three-layer systems, of which schematic image is displayed in Fig. \ref{3layerFFRHT.pdf}. 

The exchanged heat flux is 

\begin{equation}
\label{eq:heat exchange}
\begin{split}
  (2A_1+A_2)q_{123-456} &= (2A_1+A_2)\left[q_{123-4}+q_{123-5}+q_{123-6}\right]\\
  &= A_1\left[q_{14}+q_{15}+q_{16}\right] + A_2\left[q_{24}+q_{25}+q_{26}\right] + A_1\left[q_{34}+q_{35}+q_{36}\right]\\
  &= 2A_1\left[q_{14}+q_{15}+q_{16}\right] + A_2\left[2q_{24}+q_{25}\right]\\
  &= 2A_1\left[q_{1}-q_{17}\right] + A_2\left[q_{2}-q_{27}\right]{,}
  \end{split}
\end{equation}
since we assume $q_{11}=q_{12}=q_{13}=q_{21}=q_{22}=q_{23}=0$, $q_{24}=q_{26}$, and $q_i = \sum_{j=1}^{N} q_{ij}$ is satisfied. The exchanged heat flux is defined as $q_{ij}=F_{ij}\left[\sigma \left(T_i^4-T_j^4\right) + \left(\epsilon_j^{-1}-1\right)q_j - \left(\epsilon_i^{-1}-1\right)q_i\right]$, which describes $q_{17}$ and $q_{27}$ as

\begin{equation}
\begin{split}
  &q_{17} = F_{17}\left[\sigma \left(T_\text{h}^4-T_\text{amb}^4\right) + \left(\epsilon_\text{evac}^{-1}-1\right)q_7 - \left(\epsilon_1^{-1}-1\right)q_1\right]
  =F_{17}\left[\sigma \left(T_\text{h}^4-T_\text{amb}^4\right) - \left(\epsilon_1^{-1}-1\right)q_1\right]\\
  &q_{27} = F_{27}\left[\sigma \left(T_\text{h}^4-T_\text{amb}^4\right) + \left(\epsilon_\text{evac}^{-1}-1\right)q_7 - \left(\epsilon_2^{-1}-1\right)q_2\right]
  =F_{27}\left[\sigma \left(T_\text{h}^4-T_\text{amb}^4\right) - \left(\epsilon_2^{-1}-1\right)q_2\right]{.}
   \end{split}
\end{equation}
Applying these $q_{17}$ and $q_{27}$ to Equation \ref{eq:heat exchange} gives

\begin{equation}
q_{123-456} = \frac{2\left[1+F_{17}\left(\epsilon_1^{-1}-1\right)\right]q_1 + A_{21}\left[1+F_{27}\left(\epsilon_2^{-1}-1\right)\right]q_2 - \sigma\left(2F_{17}+A_{21}F_{27}\right) \left(T_\text{h}^4-T_\text{amb}^4\right)}{2+A_{21}}{.}
\end{equation}
The heat flux $q_1,2$ can be derived by solving the Equation \ref{eq:heat flux}, which is, in this case, as follows.

\begin{equation}
\label{eq:matrix_3layer}
    \left[
    \begin{array}{ccccccc}
    \epsilon_{1}^{-1} & 0 & 0 & \chi_{14} & \chi_{15} & \chi_{16} & 0\\
    0 & \epsilon_{2}^{-1} & 0 & \chi_{24} & \chi_{25} & \chi_{24} & 0\\
    0 & 0 & \epsilon_{1}^{-1} & \chi_{16} & \chi_{15} & \chi_{14} & 0\\
    \chi_{14} & \chi_{15} & \chi_{16} & \epsilon_{1}^{-1} & 0 & 0 & 0\\
    \chi_{24} & \chi_{25} & \chi_{24} & 0 & \epsilon_{2}^{-1} & 0 & 0\\
    \chi_{16} & \chi_{15} & \chi_{14} & 0 & 0 & \epsilon_{1}^{-1} & 0\\
    \chi_{71} & \chi_{72} & \chi_{71} & \chi_{71} & \chi_{72} & \chi_{71} & 1
    \end{array}
    \right]
    \left[
    \begin{array}{c}
    q_{1}\\
    q_{2}\\
    q_{3}\\
    q_{4}\\
    q_{5}\\
    q_{6}\\
    q_{7}
    \end{array}
    \right]
=\sigma
    \left[
    \begin{array}{c}
    (F_{14}+F_{15}+F_{16})\left(T_h^4-T_s^4\right) + F_{17}\left(T_\text{h}^4-T_\text{amb}^4\right)\\
    (2F_{24}+F_{25})\left(T_h^4-T_s^4\right) + F_{27}\left(T_{h}^4-T_\text{amb}^4\right)\\
    (F_{14}+F_{15}+F_{16})\left(T_h^4-T_s^4\right) + F_{17}\left(T_{h}^4-T_\text{amb}^4\right)\\
    -(F_{14}+F_{15}+F_{16})\left(T_h^4-T_s^4\right) + F_{17}\left(T_\text{s}^4-T_\text{amb}^4\right)\\
    -(2F_{24}+F_{25})\left(T_h^4-T_s^4\right) + F_{27}\left(T_\text{s}^4-T_\text{amb}^4\right)\\
    -(F_{14}+F_{15}+F_{16})\left(T_h^4-T_s^4\right) + F_{17}\left(T_\text{s}^4-T_\text{amb}^4\right)\\
    (2F_{71}+F_{72})\left\{\left(T_\text{h}^4-T_\text{amb}^4\right) + \left(T_\text{s}^4-T_\text{amb}^4\right)\right\}
    \end{array}
    \right]{,}
\end{equation}                   
Please note that $\epsilon_\text{vac}=1$, $\epsilon_{1}=\epsilon_{3}=\epsilon_{4}=\epsilon_{6}$, and $\epsilon_{2}=\epsilon_{5}$ and by assuming that $F_{ii}=0$ and $F_{12}=F_{13}=F_{23}=F_{45}=F_{46}=F_{56}=0$, the view factor matrix is 

\begin{equation}
\label{eq:matrix_viewfactor}
    \left[
    \begin{array}{ccccccc}
    F_{11} & F_{12} & F_{13} & F_{14} & F_{15} & F_{16} & F_{17} \\
    F_{21} & F_{22} & F_{23} & F_{24} & F_{25} & F_{26} & F_{27} \\
    F_{31} & F_{32} & F_{33} & F_{34} & F_{35} & F_{36} & F_{37} \\
    F_{41} & F_{42} & F_{43} & F_{44} & F_{45} & F_{46} & F_{47} \\
    F_{51} & F_{52} & F_{53} & F_{54} & F_{55} & F_{56} & F_{57} \\
    F_{61} & F_{62} & F_{63} & F_{64} & F_{65} & F_{66} & F_{67} \\
    F_{71} & F_{72} & F_{73} & F_{74} & F_{75} & F_{76} & F_{77} 
    \end{array}
    \right]
=
     \left[
    \begin{array}{ccccccc}
    0 & 0 & 0 & F_{14} & F_{15} & F_{16} & F_{17} \\
    0 & 0 & 0 & F_{24} & F_{25} & F_{24} & F_{27} \\
    0 & 0 & 0 & F_{16} & F_{15} & F_{14} & F_{17} \\
    F_{14} & F_{15} & F_{16} & 0 & 0 & 0 & F_{17} \\
    F_{24} & F_{25} & F_{24} & 0 & 0 & 0 & F_{27} \\
    F_{16} & F_{15} & F_{14} & 0 & 0 & 0 & F_{17} \\
    F_{71} & F_{72} & F_{71} & F_{71} & F_{72} & F_{71} & F_{77} 
    \end{array}
    \right]{.}
\end{equation}
Since $\sum_{j=1}^{N} F_{ij} =1$ and $A_iF_{ij}=A_jF_{ji}$ are satisfied, as well as $A_1=A_3=A_4=A_6=ab$ and $A_2=A_5=ad$, the view factors between the device surfaces and the ambient surface can be written as

\begin{equation}
\begin{split}
  &F_{17} = 1-F_{14}-F_{15}-F_{16}{,}\\
  &F_{27} = 1-F_{24}-F_{25}-F_{24}=1-2F_{24}-F_{25}{,}\\
  &F_{77} = 1-4F_{71}-2F_{72} = 1 - 4A_{17}F_{17} - 2A_{27}F_{27}{.}
\end{split}
\end{equation}
All the view factors can be expressed by $F_{14}$,$F_{15}$,$F_{16}$,$F_{24}$ and/or $F_{25}$. The view factors $F_{14}$ and $F_{25}$ can be calculated according to Equation \ref{eq:viewfactor} and other view factors can be derived by solving the view factors' algebra.

\begin{equation}
 A_2F_{24} = A_4F_{42} = A_1F_{15}
\end{equation}
\begin{equation}
\begin{split}
  (A_1+A_2)F_{12-45} &= (A_1+A_2)(F_{12-4} + F_{12-5})\\
  &= A_4 F_{4-12} + A_5 F_{5-12}\\
  &= A_4 (F_{41}+F_{42}) + A_5 (F_{51}+F_{52})\\
  &= A_1 (F_{14}+2F_{15}) + A_2 F_{25}{.}
\end{split}
\end{equation}
\begin{equation}
\begin{split}
  (2A_1 + A_2)F_{123-456} &= (2A_1 + A_2)(F_{123-4} + F_{123-5} + F_{123-6})\\
  &= A_4 F_{4-123} + A_5 F_{5-123} + A_6 F_{6-123}\\
  &= A_4 (F_{41}+F_{42}+F_{43}) + A_5 (F_{51}+F_{52}+F_{53}) + A_6 (F_{61}+F_{62}+F_{63})\\
  &= 2A_1 (F_{14}+2F_{15}+F_{16}) + A_2 F_{25}{.}
\end{split}
\end{equation}
Therefore,
\begin{equation}
\begin{split}
 &F_{24} = A_{12}F_{15}{,}\\
 &F_{15} = \frac{1}{2} \left[(1+A_{21})F_{12-45} - A_{21}F_{25} - F_{14}\right]{,}
\end{split}
\end{equation}
\begin{equation}
  F_{15} = \frac{1}{2} \left[(1+A_{21})F_{12-45} - A_{21}F_{25} - F_{14}\right]{,}
\end{equation}
\begin{equation}
  F_{16} = \left(1+\frac{A_{21}}{2}\right)F_{123-456} -  (1+A_{21})F_{12-45} - \frac{A_{21}}{2}F_{25}{.}
\end{equation}
The view factors, $F_{12-45}$ and $F_{123-456}$ can be also calculated from Equation \ref{eq:viewfactor}. 
The gap thermal conductance between three-layer system via Planck's far-field radiative heat transfer is finally,

\begin{equation}
  G_\text{FFRHT, 3-layer} = \frac{q_{123-456}}{T_\text{h}-T_\text{s}}{.}
\end{equation}

\section{5. Dispersion relation and absorption spectrum in the Si film and the three-layer system}

\begin{figure}[ht]
\centering
\includegraphics[width=\linewidth]{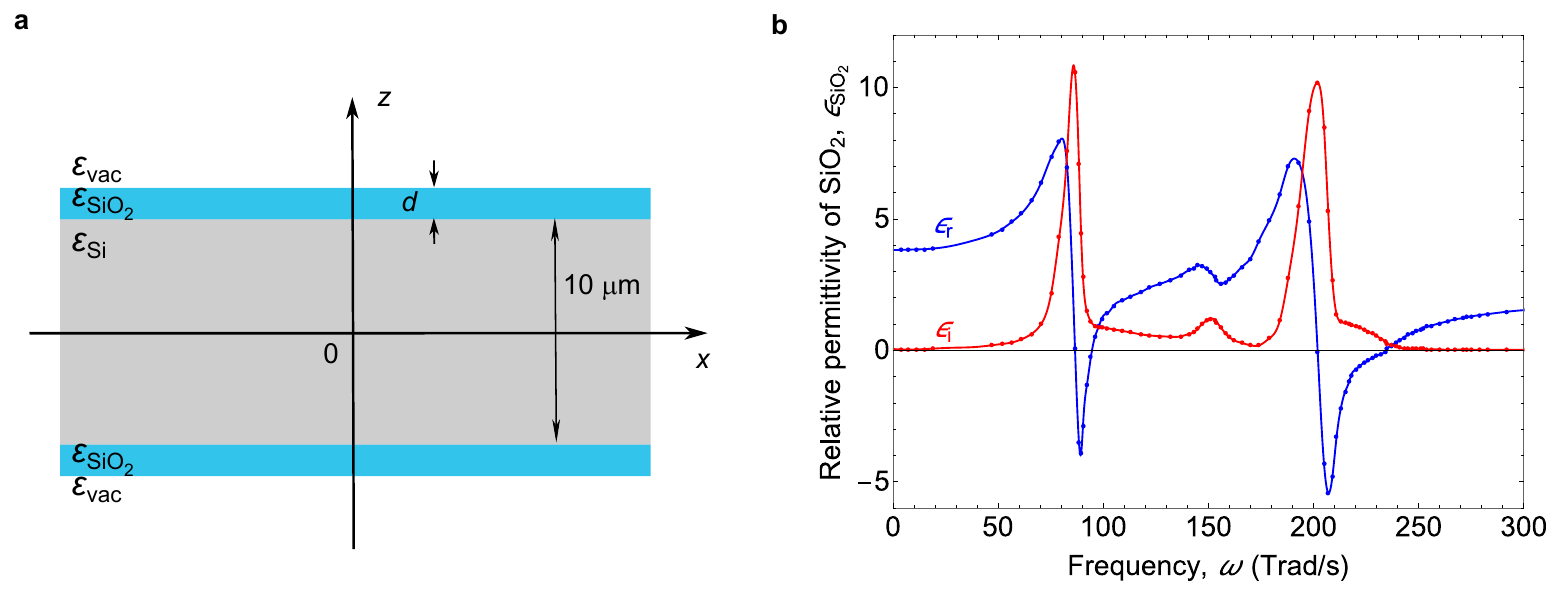}
\caption{$\mid$ \textbf{Calculations of wave vectors in the three-layer system.} \textbf{a}, Schematic of the three-layer system. The 10 {\textmu}m-thick silicon (Si) layer is sandwiched with silicon dioxide (SiO$_2$) layers. 
\textbf{b}, Real and imaginary parts of the relative permittivity of silicon dioxide \cite{palik_handbook_1998}. }
\label{Supplementary_3layer_maxwell.pdf}
\end{figure}

Figure \ref{Supplementary_3layer_maxwell.pdf}a shows the cross-section of the three-layer system in which the 10 {\textmu}m-thick silicon layer is sandwiched with the silicon dioxide layers. Silicon dioxide has a complex and frequency-dependent relative permittivity as plotted in Fig. \ref{Supplementary_3layer_maxwell.pdf}b whereas silicon has a constant and real permittivity of $\epsilon_\text{Si}=11.7$ \cite{palik_handbook_1998}. The relative permittivity in a vacuum, $\epsilon_\text{vac}$, was set as unity.
We defined the SPhP in-plane wave vector along the interfaces as a complex wave vector, $\beta = \beta_r + i\beta_i$ and the cross-plane wave vectors in each medium as $p_\text{vac}$, $p_{\text{SiO}_\text{2}}$ and $p_\text{Si}$.

\begin{figure}[ht]
\centering
\includegraphics[width=\linewidth]{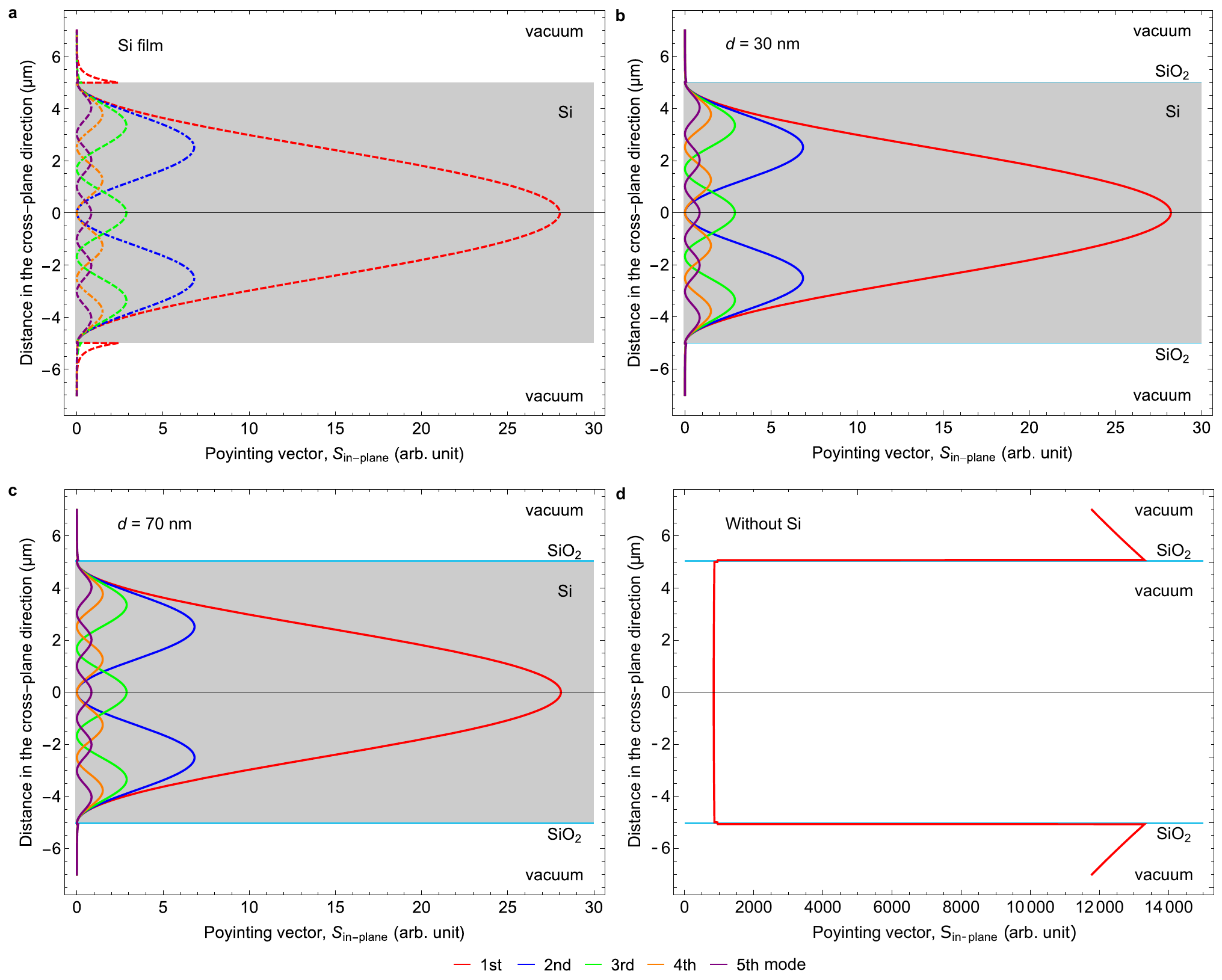}
\caption{$\mid$ \textbf{Poynting vector in the in-plane direction.} The Poynting vectors were calculated inside \textbf{a} the Si film and the three-layer system with \textbf{b} $d$ = 30 nm, \textbf{c} $d$ = 70 nm, and \textbf{d} without Si in between.}
\label{Supplementary_SPhP_Poynting.pdf}
\end{figure}

Once the Maxwell's equations with the proper boundary conditions are solved and the in-plane wave vector ($\beta = \beta _{r} + i\beta_i$) and the cross-plane wave vectors ($p_\text{vac}$, $p_{\text{SiO}_\text{2}}$, $p_\text{Si}$) are derived, elecric fields and magnetic fields can be expressed. We visualized the in-plane Poynting vectors inside the structure as shown in Fig. \ref{Supplementary_SPhP_Poynting.pdf}.

\vspace*{\stretch{1}}
\vspace*{\stretch{2}}
\newpage

\begin{figure}[ht]
\centering
\includegraphics[width=\linewidth]{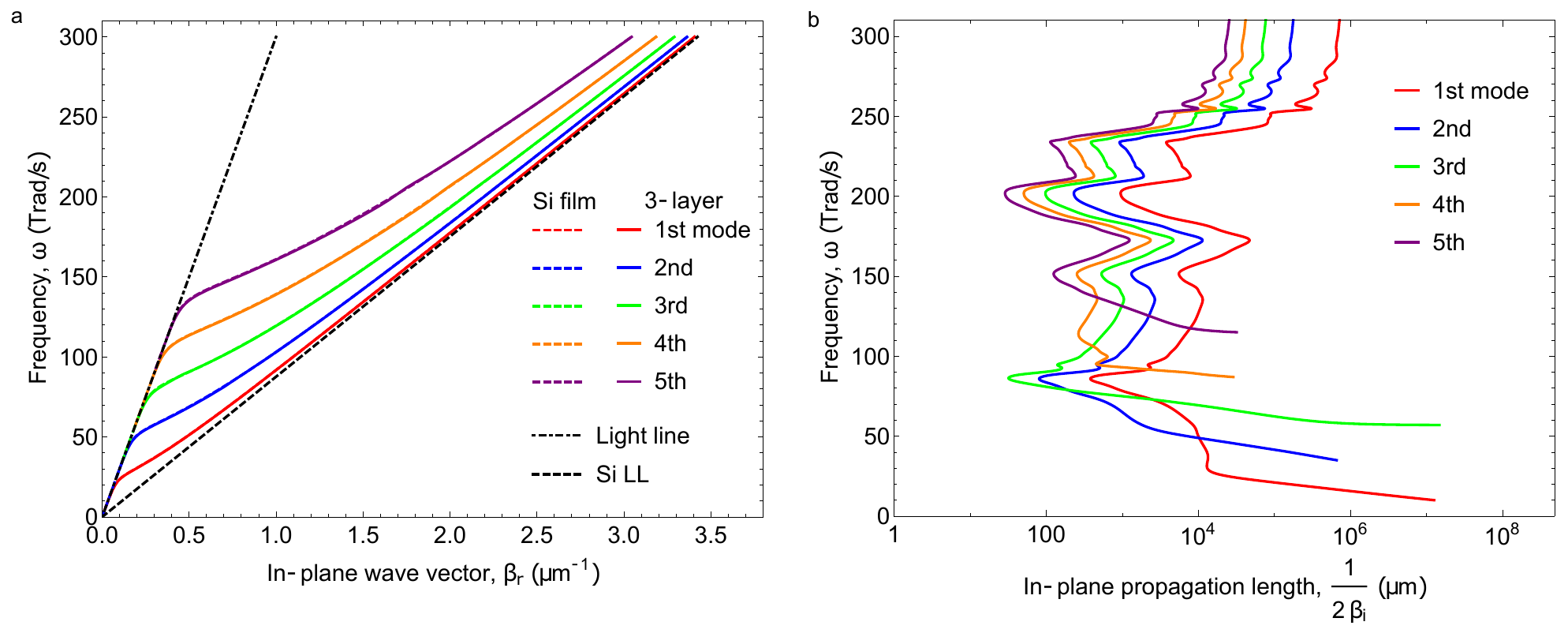}
\caption{$\mid$ \textbf{The dispersion relation and the absorption spectrum of SPhPs inside the three-layer system.} \textbf{a}, The dispersion relation of SPhPs wave vector in in-plane direction and
\textbf{b}, the SPhPs in-plane propagation length calculated for the oxide thickness of $d$ = 70 nm. 
The dashed lines in \textbf{a} are the guided resonant modes inside the 10 {\textmu}m-thick silicon film}
\label{Supplementary_SPhPdispersion.pdf}
\end{figure}

The in-plane Poynting vectors mainly exist inside Si and present guided resonant modes.
The 1st branch has the largest amplitude, contributing the most to the energy transfer.
The Poynting vector profile does not significantly differ for the Si film and the three-layer system with $d$ = 30 nm and $d$ = 70 nm.
To explain the difference in gap conductance between the Si film and the three-layer system, we calculate the dispersion relation and the propagation length spectrum in both configurations as reported in Fig. \ref{Supplementary_SPhPdispersion.pdf}.

In Fig. \ref{Supplementary_SPhPdispersion.pdf}a, the coloured solid lines are the first to fifth modes in the three-layer system, and the dashed lines are the modes in the 10 {\textmu}m-thick Si film.
Figure \ref{Supplementary_SPhPdispersion.pdf}a shows that the dispersion curves of the three-layer system match the ones of the silicon film.
This correspondence and the predictions shown in Fig. \ref{Supplementary_SPhP_Poynting.pdf} confirm the hybridization of SPhP with guided resonant modes of Si.
However, these guided resonant modes cannot be excited thermally without the SiO$_2$ nanolayers, which generate SPhPs propagating along the interfaces.
The guided resonant modes can only be thermally excited if absorption is large enough to couple heat to the electromagnetic field.
Although the dispersion relation of the three-layer system and of the pristine Si are mostly overlapped, the propagation length spectrum in the three-layer system reported in Fig. \ref{Supplementary_SPhPdispersion.pdf}b reveals the basic difference between the pristine Si film and the three-layer system. Whereas the Si film has negligible absorption (dielectric function with imaginary part equals to zero) thus infinite propagation length, the three-layer system absorbs significantly and supports the generation and propagation of SPhP waveguide modes in the full spectrum.
Consequently, those modes contribute to the gap conductance only when they are thermally excited in the SiO$_2$ nanolayers while remaining unexcited in the case of the non-absorbing pristine Si film.
Taking into account that the SPhP propagation length is greater than the length of the three-layer system, as shown in Fig. \ref{Supplementary_SPhPdispersion.pdf}b, the device size remains within the range where the SPhP transport regime is ballistic.

\section{6. The SPhP propagation with/without Si layer}

In this section, we show that the generation of SPhP resonant guided modes is indeed due to the presence of the Si layer.
Figure \ref{Supplementary_SPhP_Poynting.pdf}c and \ref{Supplementary_SPhP_Poynting.pdf}d show the comparison of the SPhP in-plane Poynting vector with and without Si layer, respectively.
In the presence of the Si layer, Fig. \ref{Supplementary_SPhP_Poynting.pdf}c highlights that SPhP resonant guided modes are excited, and the electromagnetic flux is mainly distributed inside the non-absorbent Si layer, as discussed in the main manuscript and the Supplementary Note 6.
By contrast, when the Si layer is replaced by vacuum, SPhPs propagate mainly along the SiO$_2$ nanolayers and there is no SPhP resonant guided modes.
The presence of the Si layer is therefore essential for the hybridization of SPhPs through these resonant guided modes.

\section{7. The influence of Pt thermometer on the in-plane Poynting vector}

\begin{figure}[ht]
\centering
\includegraphics[width=\linewidth]{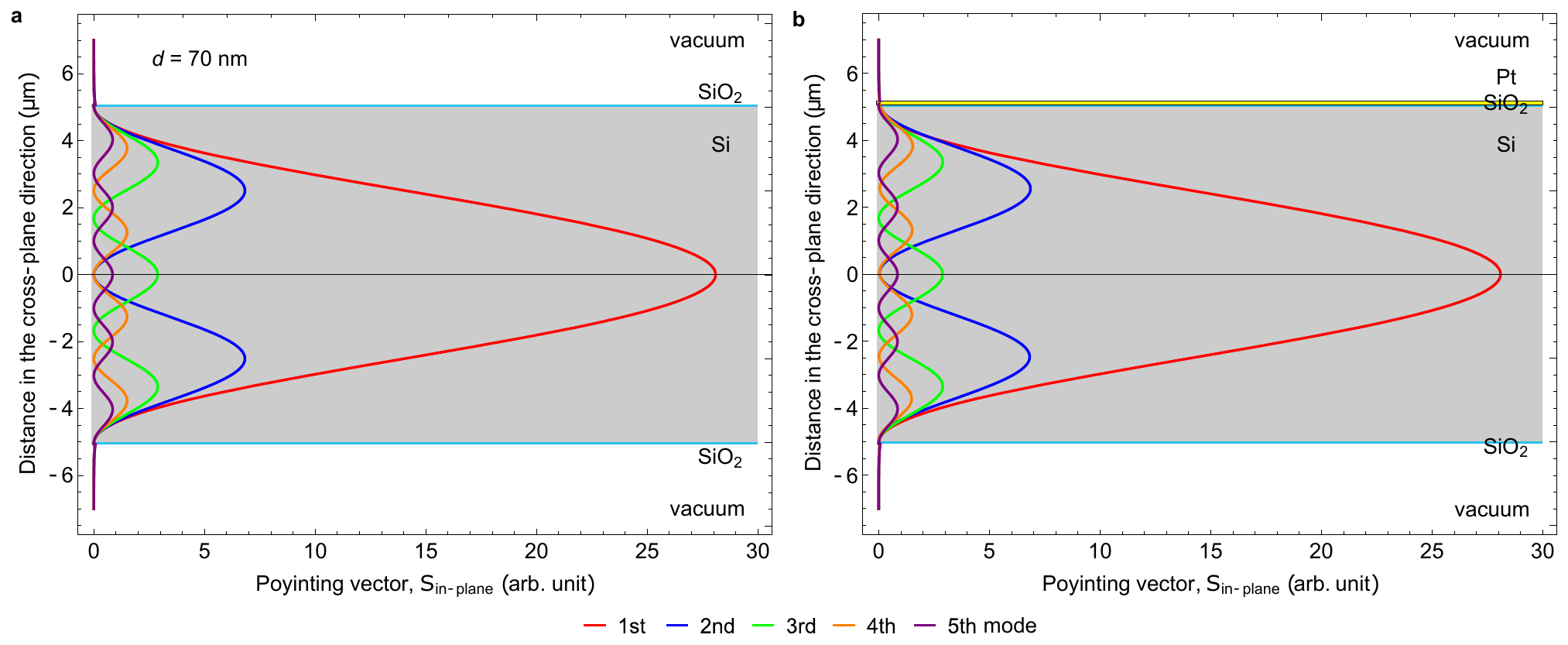}
\caption{$\mid$ \textbf{The influence of the Pt layer on Poynting vectors inside the three-layer system.} Poynting vector in the in-plane direction inside the three-layer system \textbf{a} without and \textbf{b} with the Pt layer of 100 nm thickness}
\label{Supplementary_SPhP_Poynting_Pt.pdf}
\end{figure}

We solved Maxwell's equation in the presence of the Pt layer of 100 nm thickness and calculated the in-plane Poynting vector, as shown in Fig. \ref{Supplementary_SPhP_Poynting_Pt.pdf}. Although the energy at the interface between the top SiO$_2$ and the Si layers are slightly higher due to the reflection at the Pt layer, the in-plane Poynting vectors with and without the Pt layer do not show the substantial difference. Thus, we confirmed that the influence of the Pt thermometer on the energy inside the three-layer structure is negligible.

\section{8. Gap thermal conductance in the three-layer system}

To evaluate the gap conductance in the three-layer system, we consider, firstly, the contribution of Planck's radiation that was proven in the case of the pristine Si film.
As explained in Supplementary Note 6, a second contribution should be added to the gap conductance to take into account in-plane guided resonant modes that are thermally activated only in the case of the three-layer system.

To derive this second contribution, we start with the definition of the heat flux as established by the classical theory of far-field radiation where only the in-plane azimuthal angle $\phi$ is taken into account, as

\begin{equation}
\label{eq:heatflux}
  dq = \frac{dE}{dydzdt}= \frac{dE}{dy b dt}{,} dE = \hbar \omega f(\omega) D(\omega) d\omega d\phi
\end{equation}
where $b$ is the total thickness of the plate, $\hbar$ is the Planck's constant, $\omega$ the angular frequencies, $f(\omega)$ the Bose-Einstein distribution function and $D(\omega)$ the two-dimensional (2D) density of states, given by

\begin{equation}
\label{eq:DOS}
  D(\omega) d\omega d\phi = \frac{dxdydp^2}{h^2}= \frac{dxdydk^2}{(2\pi)^2} =\frac{dxdy}{(2\pi)^2} dk(kd\phi) = \frac{dxdy}{(2\pi c)^2}\omega d\omega d\phi{,}
\end{equation}
where $p = \hbar k$ and $k=\frac{\omega}{c}$. By combining Eq. \ref{eq:DOS} and Eq. \ref{eq:heatflux}, the differential of the heat flux can be rewritten as

\begin{equation}
\begin{split}
  dq =  \frac{dE}{dy b dx}\frac{dx}{dt} = \frac{dE}{dxdy b }c\cos{\phi} = \frac{\hbar \omega c}{b} f(\omega) \frac{\omega}{(2\pi c)^2} d\omega \cos{\phi} d\phi{.}
 \end{split}
\end{equation}
Integration of the above equation yields the heat flux as

\begin{equation}
\begin{split}
  q = \int^{\infty}_0 \frac{\hbar c}{(2\pi c)^2 b} \omega^2 f(\omega) d\omega \int^{\pi/2}_{-\pi/2} \cos{\phi} d\phi = \frac{\sigma_\text{2D}}{b}T^3{,}
 \end{split}
\end{equation}
where $\sigma_\text{2D} = 4\zeta (3)\frac{k_B^3}{ch^2}$ W m$^{-1}$K$^{-3}$. Therefore, the gap thermal conductance $G_\text{SPhP}$ in the in-plane direction is

\begin{equation}
\label{eq:G_SPhP}
    G_\text{SPhP}=\frac{q}{T_\text{h}-T_\text{s}}=\frac{\epsilon \sigma_\text{2D} a (T_\text{h}^3 - T_\text{s}^3)}{T_\text{h}-T_\text{s}}{,}
\end{equation}
where $\epsilon = \frac{\epsilon_\text{3-layer}}{F^{-1}+1-\epsilon_\text{3-layer}}$ is the standard effective thermal emissivity between two radiating flat surfaces, $a$ is the width of the plate (see Fig. \ref{3layerFFRHT.pdf}), $F$ refers to the view factor and $\epsilon_\text{3-layer}$ to the emissivity of the three-layer system.
Equation \ref{eq:G_SPhP} represents our 2D model for the SPhP contribution to the gap thermal conductance.

\section{9. Comparison of our model with the previous experimental data}

To confirm the validity of our model of $G_\text{SPhP}$ not only for macroscale structures, but also to nanoscales ones, we estimated the gap thermal conductance reported in Ref \cite{thompson2018hundred} using Eq. \ref{eq:G_SPhP}.
Our model is based on the in-plane propagation of SPhPs contributing to the gap thermal conductance on top of the conventional Planck's radiative heat transfer.

\begin{figure}[ht]
\centering
\includegraphics[width=\linewidth]{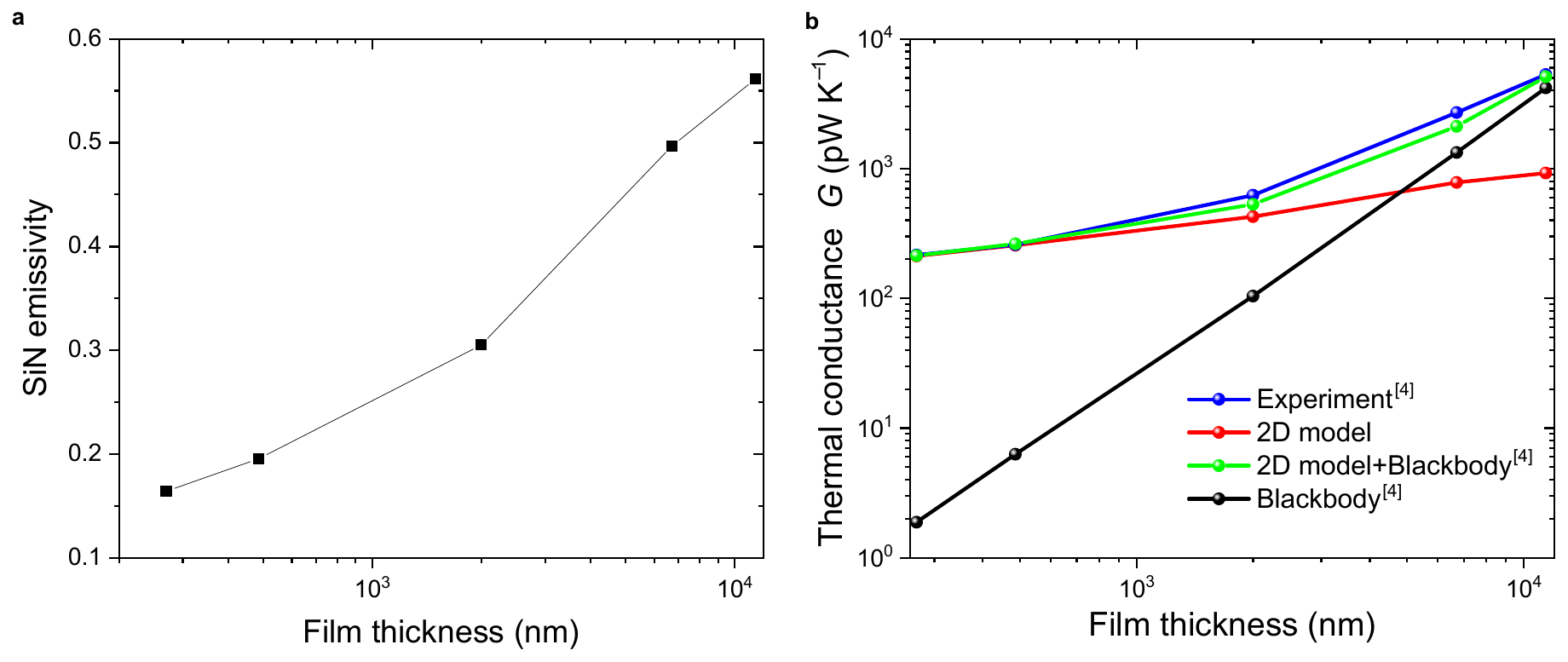}
\caption{$\mid$ \textbf{Comparison of our model with the previous work \cite{thompson2018hundred}.} \textbf{a} The emissivity of SiN reported in Ref\cite{palik_handbook_1998}.
\textbf{b} Our 2D model is in comparison with the experimental results of Ref \cite{thompson2018hundred}. 
The view factor of 1 is used to calculate the 2D gap conductance.
The nanofilm width is 60 {\textmu}m.}
\label{2DPramod}
\end{figure}

Taking into account that the experimental results reported in Ref \cite{thompson2018hundred} were obtained for far-field radiation between two silicon nitrite (SiN) nanofilms, our 2D model in Eq. \ref{eq:G_SPhP} has been applied with a 2D view factor equals to unity, a width $a$ = 60 {\textmu}m \cite{thompson2018hundred}, and the SiN emissivity shown in Fig. \ref{2DPramod}(a) \cite{palik_handbook_1998}.
The comparison between our theoretical results and the literature experimental data is shown in Fig. \ref{2DPramod}(b).
Note that the proposed 2D model together with the blackbody radiation describes well the experimental data for different membrane thicknesses.
This agreement confirms the applicability of our 2D model to explain the super Planckian thermal transport driven by SPhPs. 

\newpage
\section{10. Temperature distribution on the hot plate}

\begin{figure}[ht]
\centering
\includegraphics[width=\linewidth]{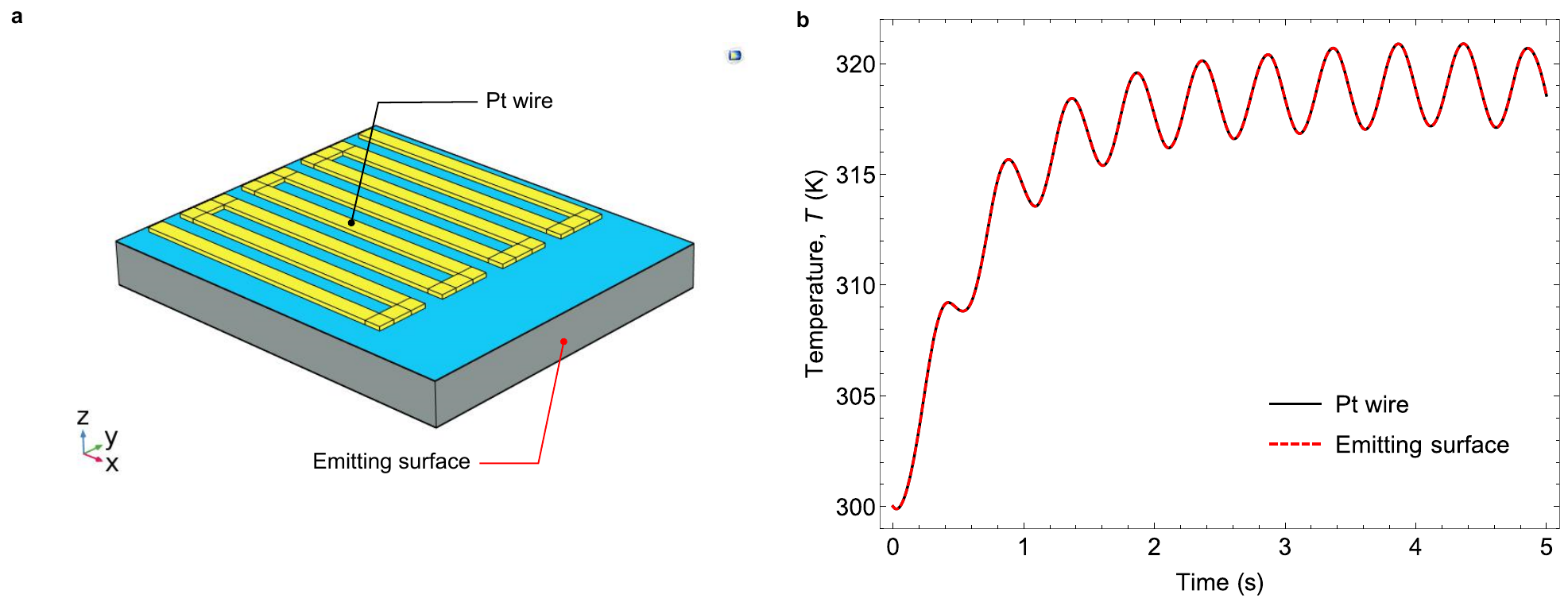}
\caption{$\mid$ \textbf{The estimations of temperature distribution on the hot plate.} \textbf{a}, Schematic of the computational model. The 10 {\textmu}m-thick Si layer (in grey) is sandwiched with SiO$_2$ layers (in blue) of 70 nm in thickness. The temperature on the Pt wire and on the emitting surface are computed.
\textbf{b}, Temperatures on the Pt wire and on the emitting surface.}
\label{Supplementary_Jouleheating.pdf}
\end{figure}

To confirm that the temperature measured on the platinum (Pt) thermometer is equivalent to the temperature of the emitting surface of the hot plate, we determined the temperature distribution over the hot plate by using FDTD simulation.
The computational system consists of a 10 {\textmu}m-thick Si sandwiched with the SiO$_2$ layers (70 nm thick), as shown in Fig. \ref{Supplementary_Jouleheating.pdf}a.
By using a sinusoidal current with an oscillating frequency of 1 Hz to generate Joule heating, the time evolution of the temperature of the Pt wire and the emitting surface are calculated and shown in Fig. \ref{Supplementary_Jouleheating.pdf}b.
One can see that both spots are at the same temperature, which indicates that the temperature measured by the Pt thermometer represents an accurate estimation of the emitting surface temperature.
This result is expected to apply for all modulation frequencies $f$= 1--100 Hz used in our experiments.
As for all of the frequencies, the propagation (diffusion) length $\mu = \sqrt{\alpha/ 2\pi f}$ of the excited thermal waves is greater than the distance between the Pt wire and the emitting surface (20 {\textmu}m).
For instance, as the system consists of Si mainly, its thermal diffusivity $\alpha \approx$ 80 mm$^2$⁄s, $\mu$ =12732.4 {\textmu}m (127.3 {\textmu}m) for $f$=1 Hz ($f$=100 Hz) and therefore the thermal waves travel from the Pt wire to the emitting surface with practically no attenuation.
This is the reason why these two spots exhibit nearly the same temperature field. 

\section{11. Deviations from planarity of the device}

\begin{figure}[ht]
\centering
\includegraphics[width=0.6\linewidth]{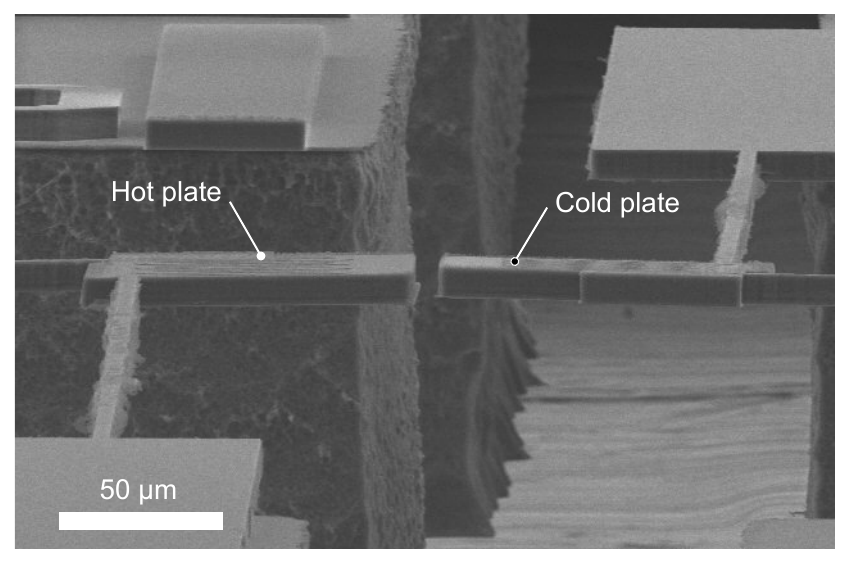}
\caption{| \textbf{Deviation from planarity of the devices}. SEM image of the device from the side view. The devices were flat and well-aligned.}
\label{Supplementary_flat_device.pdf}
\end{figure}

Our samples were confirmed to be flat and well-aligned, as shown by SEM observations (Fig. \ref{Supplementary_flat_device.pdf}).
The support beams have the width, length and thickness of 4, 706, and 10 {\textmu}m, respectively.
The width being smaller than the thickness prevents buckling in the out-of-plane direction.
The two beams supporting the plate are aligned by 90 degrees to prevent both buckling in the in-plane direction and radiative heat exchange between the beams of the hot and the cold plates.

\newpage
\bibliography{Reference.bib}